\documentclass{article}

\usepackage{PRIMEarxiv}

\usepackage[utf8]{inputenc} 
\usepackage[T1]{fontenc}    
\usepackage{hyperref}       
\usepackage{url}            
\usepackage{booktabs}       
\usepackage{amsfonts}       
\usepackage{nicefrac}       
\usepackage{microtype}      
\usepackage{lipsum}
\usepackage{fancyhdr}       
\usepackage{graphicx}       
\graphicspath{{media/}}     
\usepackage{amsmath}
\usepackage{caption}
\usepackage{subcaption}

\usepackage{makecell}
\usepackage{multirow}

\title{Large Language Models for Social Networks: Applications, Challenges, and Solutions
}

\author{
  Jingying Zeng, Richard Huang, Waleed Malik, Langxuan Yin, Bojan Babic, \\ \textbf{Danny Shacham}, \textbf{Xiao Yan}, \textbf{Jaewon Yang}, \textbf{Qi He}\\
  \vspace{0.01mm}\\
  Nextdoor, USA\\
}

\begin{document}
\maketitle

\begin{abstract}
Large Language Models (LLMs) are transforming the way people generate, explore, and engage with content. We study how we can develop LLM applications for online social networks. Despite LLMs' successes in other domains, it is challenging to develop LLM-based products for social networks for numerous reasons, and it has been relatively under-reported in the research community.

We categorize LLM applications for social networks into three categories. First is knowledge tasks where users want to find new knowledge and information, such as search and question-answering. Second is entertainment tasks where users want to consume interesting content, such as getting entertaining notification content. Third is foundational tasks that need to be done to moderate and operate the social networks, such as content annotation and LLM monitoring. For each task, we share the challenges we found, solutions we developed and lessons we learned. To the best of our knowledge, this is the first comprehensive paper about developing LLM applications for social networks.

\end{abstract}

\keywords{Generative AI \and Large Language Models \and Social Networks \and Machine Learning \and Reinforcement Learning \and Rejection Sampling \and Natural Language Generation}

\newpage

\tableofcontents

\newpage
\section{Introduction}

Large Language Models(LLMs), such as GPT-3 \cite{brown2020language} and LLAMA \cite{touvron2023llama}, are AI models that can perform tasks by reading instructions. These models are pre-trained on large amounts of text corpora and then fine-tuned to perform tasks by following the instructions~\cite{ouyang2022training}. LLMs have shown impressive performance across different Natural Language Processing (NLP) tasks, and shown a great potential to adapt to unseen tasks by in-context few-shot learning \cite{brown2020language, xie2021explanation}. If we want to customize LLMs for a specific task, we can fine-tune LLMs with labeled data in the same way we train other AI models. LLMs bring a lot of opportunities for new AI applications due to its versatility.

In this paper, we aim to develop LLM applications for online social networks.
Online social networks are the websites where people can connect and interact to each other for different purposes. People use Facebook and Instagram to interact with their friends, LinkedIn to connect with their colleagues, and Nextdoor to learn what is going on in their neighborhoods. Social network sites have been developed (and incentivized) to maximize the amount of human-to-human interactions rather than human-to-computer interactions such as human-to-LLMs. It is fascinating to imagine how LLMs can interact with users to help them interact with other users.

We note that different social network users have different reasons for using social networks, and thus decide to define a few use case types for LLMs and tackle each type separately. Although the primary reason may be social interactions with other users, the goal of interactions can be different. Sometimes users want to get information or new knowledge. For example, users would post questions like ``is this restaurant good?''. Or, users want to have fun by browsing interesting content, like watching short video clips about cats and dogs. Lastly, there are foundational tasks that need to be done regardless of applications, such as monitoring the health of LLMs, flagging malicious content on the site and so on. In summary, we categorize the LLM applications for social networks as follows:
\begin{itemize}
    \item \textbf{Knowledge tasks}: Tasks where users want to get new knowledge or new information, such as Search among social posts, Asking questions to other social network users;
    \item \textbf{Engagement tasks}: Tasks where we use LLMs to increase user engagement, e.g. creating interesting content for notifications;
    \item \textbf{Foundation tasks}: This includes tasks that bring impacts across many applications horizontally. For example, how to build to manage the API usage and LLM health, and how to build knowledge graph with LLMs belong to this type.
\end{itemize}


Developing LLM applications for social networks is a challenging job for many reasons. First, existing LLMs are not trained for the tasks we mentioned above. For the knowledge tasks, social network users want to know about knowledge specific to the user's social context, such as ``which park around my home has a nice tennis court'', and the training corpora for LLMs do not contain this kind of local, long-tail knowledge. For the engagement tasks, it is even trickier to train LLMs because whether the content is engaging to users is highly subjective and specific to each social network. For example, content popular among Instagram is very different from content popular among Twitter (or X) users.

Second challenge is that social networks are evolving over time while LLMs are static. For knowledge tasks, the knowledge and information keeps changing in the social networks, but LLMs' knowledge does not get updated after training. For engagement tasks, users' tastes changes dynamically over time. 

Third challenge is evaluation: to evaluate LLMs' outputs and remove hallucinations (mentioning something that is incorrect). LLMs' outputs are free form text which is time consuming to evaluate for human annotators.

In this paper, we discuss how to tackle the above challenges in a systematic way. These solutions come from 1-year experience in developing several LLM applications for Nextdoor, a hyper-local social network for neighborhoods. We found that different task types pose different challenges and need different solutions. Therefore, we will disucss challenges and solutions for each type in the following sections. We believe that, although we present challenges and solutions in the context of Nextdoor, these lessons are applicable to other social network settings.

Here is a summary of solutions we discuss in the paper. For knowledge tasks, we employ Retrieval Augmented Generation (RAG) to retrieve the posts and comments from the social network that are relevant to a given task or question. RAG addresses the first and second challenges by providing LLMs with up-to-date knowledge from the social networks. Even with RAG, we found that the underlying LLMs make mistakes by referencing wrong content. To address this, we fine-tune LLMs by teaching which types of answers should be preferred. We employed a fine-tuning method that is much simpler than Reinforcement Learning from Human Feedback~\cite{ouyang2022training}. For the evaluation challenge, we adopted a LLM-based approach to evaluate the quality of LLM outputs holistically.

For the engagement tasks, we developed a reward model that predicts if the users would interact with the content made by LLMs. We trained a reward model with our own user engagement data because each social network's users have different tastes. This reward model addresses the first challenge. To address the second challenge, we retrain the reward model periodically. For the engagement tasks, there is no evaluation challenge because we can use user engagement metrics.

For the foundation tasks, the above challenges do not apply. Foundation tasks are not domain specific nor time sensitive. Evaluation is also relatively straightforward. For the foundation tasks, we discuss how we optimize the cost of LLMs serving and how to monitor LLM performance efficiently.


Our contributions are the following:
\begin{itemize}
\item We give a high level overview on a broad range of LLMs applications for social networks and discuss all possible challenges one would face.
\item We share how we developed the solutions and how we productionized them end-to-end. Our solutions are robust and scalable enough for real-world large social networks. To our knowledge, this is the first paper to share all the details in productionizing LLMs for social network applications.
\end{itemize}

\newpage
\section{Knowledge Tasks}
Knowledge tasks are fundamental in social networks since lots of users come to social network platforms for information seeking. In this chapter, we will describe the LocalGPT system, a neighborhood-specific LLM that we build for domain-specific question-answering.

\subsection{Introduction to LocalGPT System}
\begin{figure}[h]
    \centering
    \includegraphics[width=1\textwidth]{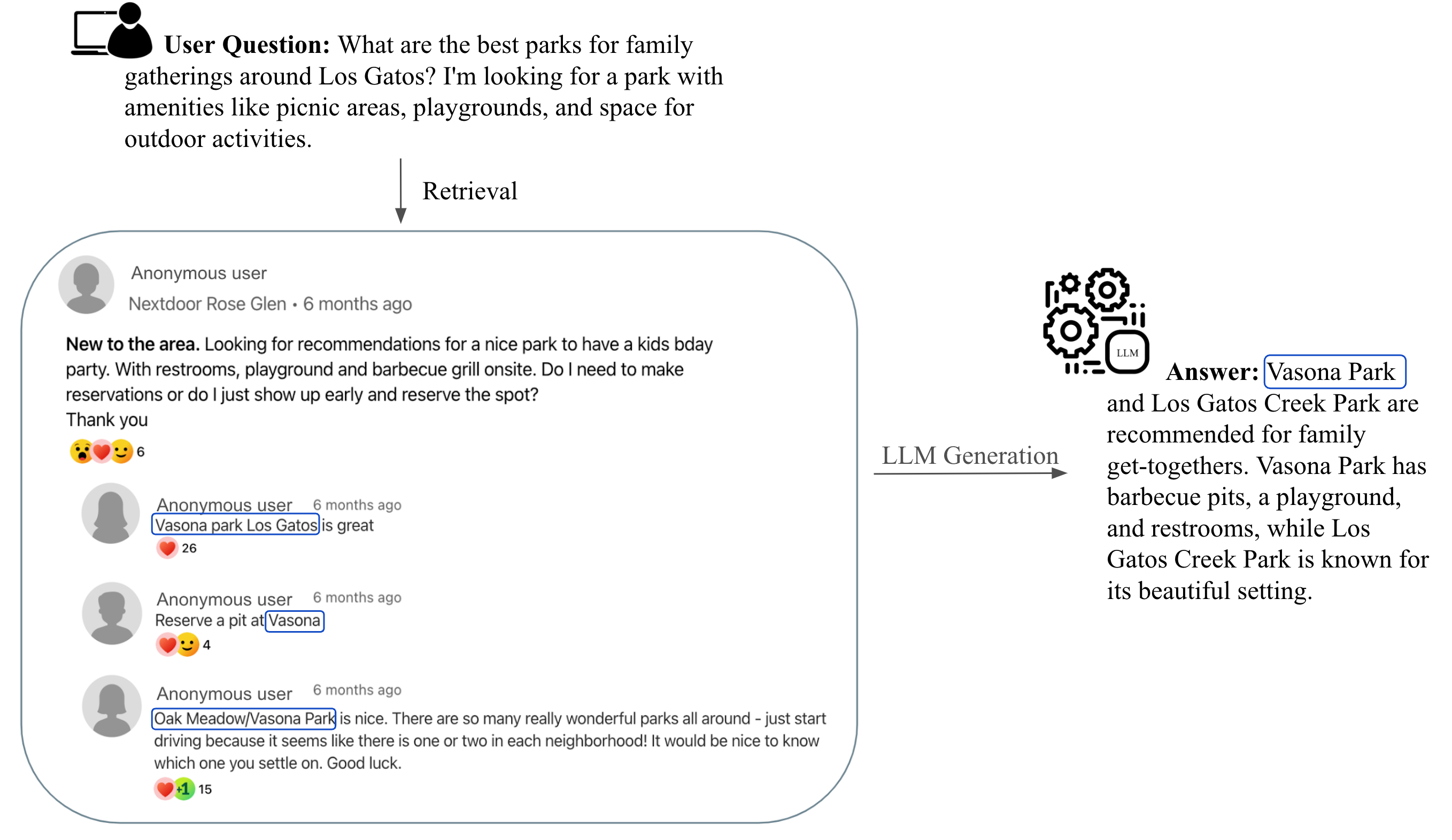}
    \caption{An Illustration of the LocalGPT System.}
    \label{example}
\end{figure}

We build LocalGPT, a LLM that contains the local knowledge from the Nextdoor data. LocalGPT can answer users' questions and search queries about users' neighborhoods like Figure~\ref{example}. LocalGPT can be used in search, like Microsoft BingGPT. LocalGPT may be used to add a comment to a user's post that has not been answered by other users.

We note that the existing LLMs like ChatGPT have shortcomings for our purposes because these models do not have up-to-date information due to the cutoff date in the training data, and because they are not trained with questions specialized to Nextdoor's domain \cite{vu2023freshllms}.
For instance, in our use case, to provide answers to the neighborhood-specific questions such as ``Does anyone know of a reputable and affordable wedding photographer in the San Francisco Bay Area?'' or ``Do you have any real-time updates on the status of wildfires in Sonoma county'', pre-trained LLMs or traditional web search might fail or provide a generic answer due to insufficient domain-specific knowledge stored or no real-time information about the topic. 

To address the limitation of existing LLMs, our LocalGPT employs two solutions. First, we apply Retrieval-Augmented-Generation (RAG): given a user query, we retrieve the most relevant documents in the Nextdoor data, and use the documents to generate an answer. Second, we fine-tune LLMs with our data. In particular, we fine-tune LLMs so that they learn what questions our users would ask (instruction tuning), and also what kinds of answers the users would prefer (preference learning). Our final models would combine these two approaches. Figure \ref{example} is an end-to-end illustration of the LocalGPT system. 

\textbf{Retrieval-Augmented-Generation}: Our first method is RAG.
In RAG, we retrieve relevant Nextdoor documents (user's posts and comments) for a user query. Then we ask pre-trained LLMs to generate answers using the documents~\cite{ram2023context}. RAG provides multiple benefits. First, RAG allows us to answer questions with the up-to-date information that happened after the cutoff date of the LLM training data. Second, RAG reduces hallucination because we ask LLMs to provide answers within the retrieved documents. 

On the other hand, using RAG only has limitations. In some cases, the retrieved knowledge can be partially or completely irrelevant to the given question, which makes it challenging for the pre-trained LLM to selectively use the information in the provided context during answer generation without training. This issue is magnified especially when we use noisy posts and associated user comments of all kinds from social network platforms to build indexes. Additionally, for time-sensitive questions, the up-to-date knowledge retrieved and provided as in-context might contradict with the pre-trained LLM's prior beliefs, resulting in the model generating answers based on the out-dated knowledge rather than attending to the new factual information in the given context \cite{meng2022locating, hernandez2023inspecting, zhang2023tell}.


\textbf{Fine-tuning:} To address the limitation of RAG, we fine-tune LLMs with the instructions related to Nextdoor's use cases (Instruction tuning). Our goals are the following: First, fine-tuned LLM can generate answers with higher quality than pre-trained LLM in terms of relevance, engagingness, coherence and readability. Second, the fine-tuned LLMs should provide useful answers even in corner cases, such as no documents retrieved or conflicting information retrieved. Third, we use Supervised fine-tuning (SFT) instead of Reinforcement learning (RL) because production-scale APIs only support SFT. It is a challenge to satisfy all these objectives and corner cases without using RL~\cite{touvron2023llama}.

To achieve the above goals, we design a novel training framework, knowledge infection training. The knowledge injection training is a multi-objective optimization to fine-tune LLM using tasks appearing in pre-training for downstream long-form Q\&A. During the fine-tuning phase, we utilize a long-form Q\&A task for knowledge injection. We also tailor the conventional multiple choice question answering (MCQA) task as knowledge preference data to upweight the token probability of knowledge-preferred responses, where we prefer a response generated using knowledge from our local knowledge bank rather than the parametric knowledge learned from pre-training. Additionally, we instruction fine-tune with RAG (IRAG) to instruct model to provide an answer by extracting factually relevant information from the provided context and ignore the spurious correlations. We empirically observe that IRAG not only benefits the in-distribution questions, but also enhances the performance and significantly reduces hallucination on out-of-domain questions.


Our knowledge injection training is a reinforcement-learning-free approach. The combination of tasks we use in gradient updates creates a more deliberate attention to allow the fine-tuned LM to selectively extract and utilize the factually related information contained in the given context for better generalization. With this training framework, we effectively inject neighborhood-specific knowledge to off-the-shelf LMs, which mitigates hallucination during answer generation. Our fine-tuning strategy bypasses training a reward model and utilizing reinforcement learning (RL) techniques, providing a more direct way to adapt domain-specific knowledge and lower the factual errors in downstream knowledge-rich tasks such as domain-specific Q\&A.  

To accelerate the training process and reduce the costs from human evaluation, we develop a scalable automatic evaluation system to comprehensively evaluate the truthfulness and contextual coherence of the responses, and monitor the negative effects that are potentially introduced by fine-tuning. This evaluation system overcomes the challenge of the nonexistence of a benchmark dataset for our domain-specific Q\&A in real-world distribution, and can be generalized to any factoid Q\&A tasks evaluation. 

As a real-world Q\&A application, our system design provides fundamental supports on the freshness of the knowledge base for the models to perform in-context learning. In the retrieval system, the vector store is updated in real-time on the creation of posts, and separated from model serving. By incorporating up-to-date knowledge retrieved from knowledge source for conditioning generation, the LocalGPT system is able to generate more reliable answers for domain-specific and time-sensitive questions.

The key contributions of this work are as follows:
\begin{itemize}
    \item We build a LocalGPT system, a real-world application for domain-specific Q\&A, that utilizes both fine-tuning to endow pre-trained LLM with domain-specific knowledge and RAG architecture to enhance out-of-the-box capabilities. The system design decouples the model serving from vector store updates to provide the fine-tuned model with always-fresh information to answer time-sensitive questions and dynamically adapt out-dated ground truth answers from knowledge injection training in real-time.

    \item We develop a multi-objective knowledge injection training framework to adapt off-the-shelf closed-source LLMs to domain-specific Q\&A task. We propose a straightforward reward-free approach to inject knowledge and improve grounding using preference data and instruction-tuning in the training phase, which significantly improves generalization ability and reduces hallucination during answer generation. 
    
    \item For knowledge injection training, our work sheds light on the effectiveness of i) using tasks similar to pre-training data to adapt the model to specialized downstream knowledge-rich tasks ii) using preference data, instruction-tuning, and in-context learning to further align the LM's attention focus with our knowledge preference and improve knowledge grounding. The ablation studies show the effectiveness of each component in the training framework.

\end{itemize}

In the following sections, we will describe the details of the RAG that integrates a language model with retrieval system without further training in section \ref{sec_base_rag}, the details of multi-objective knowledge injection training framework to further adapt pre-trained LLM to domain-specific Q\&A in section \ref{sec_multi_objective_kit}, the automatic evaluation system in section \ref{eval_system}, and the experiments results in \ref{experiment}.

\subsection{Base Retrieval-Augmented Generation (RAG)} \label{sec_base_rag}
This section describes how we build LocalGPT \emph{without fine-tuning LLMs}. In particular, we discuss using Retrieval-Augmented-Generation (RAG). Given a user's question, we retrieve relevant posts and comments and use LLMs to generate an answer using the posts and comments. For model training, we will discuss it in Section~\ref{sec_multi_objective_kit}. 

LLMs pre-trained on public data alone does not capture neighborhood-specific knowledge as our internal data does. However, without further training, off-the-shelf LLMs can be integrated and augmented with internally built retrieval systems for domain-specific question-answering \cite{ram2023context, lewis2020retrieval}. In-context RAG model is an industry standard approach, similar to that has been used in Microsoft Bing, and other applications. The key advantages of this approach are that it is a fast-built, relatively cheaper, and stable approach. 

\textit{Our base RAG leverages  the retrieval system with the generative power of LM. It is built based on integrating a pre-trained LLM with our internal retrieval system without any specialized training on the language model}. It consists of two sub-modules: a searched-based knowledge-retriever module that fetches the top-K of the most relevant posts and comments via embedding-based search to provide source knowledge, and a retrieval-augmented language generator module that leverages LLMs' parametric knowledge with the retrieved contexts provided at inference time to produce final answers back to users. Some research shows that the conflicts between parametric knowledge stored in LMs and new evidence in retrieved documents affect the answer generation \cite{chen2022rich, liu2023lost}. To alleviate this problem, we prompt LM to perform a summarization task at inference time given the relevant documents provided. For opinion-wise questions, without explicitly asking the LM to summarize based on the given context, it may encourage the freedom of the generated answers, which can potentially lead to some undesirable aspects. The base RAG serves as the baseline of our LocalGPT system. With reference posts, it is easier for a human to evaluate and validate the generated answers. 


\subsubsection{Implementation Details}
Building Base RAG consists of three steps: data generation using the most up-to-date posts and comments, index building through OpenAI embeddings, and connecting the retrieval system with a language generator model. 

\textbf{I. Data generation for Base RAG}: We first collect the most recent posts and comments through our Nextdoor's internal database. To make the collected data a high quality and trustworthy resource for index building, we select the posts with comments and  reactions above certain thresholds. Additionally, we prepare each document as a post concatenated with one associated thread. Each thread contains a parent post and a single comment section with respective replies in the same thread. In this way, for each post, we are able to capture different conversation branches in the comment section. After preprocessing, there are approximately 300K posts, branching into 1M threads for further index building.

\begin{figure}[h]
    \centering
    \includegraphics[width=0.35\textwidth]{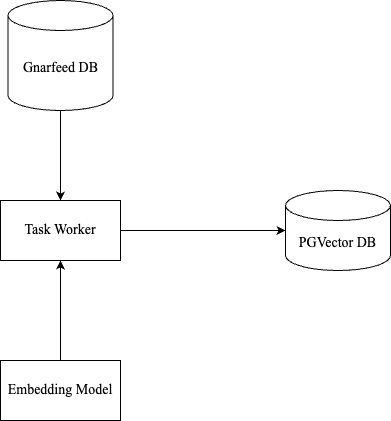}
    \caption{System Diagram of Index Building.}
    \label{index_building}
\end{figure}

\textbf{II. Index building}: In this retrieval system, the search index is built based on OpenAI embeddings. The advantage of using OpenAI embeddings is because of its long (3k+) context window. Other embedding models (e.g 512 token windows) are considered relatively too short to capture the entire posts and comments. In each Designated Market Area (DMA), we create a vector embedding for every post. The user queries are embedded in the same vector space such that Approximate Nearest Neighbors (ANN) can be used to retrieved the relevant posts.

Figure \ref{index_building} is a diagram illustrating how indexes are built. We store all the posts and comments in a Postgres database named Gnarfeed. At the time a post is created, the Task Worker calls the embedding model and stores the resulting embedding and post ID into the vector store, where Task Worker is Nextdoor’s internal task queue system. The vector store is built by Postgres. Postgres contains PGVector to support retrieval for related posts.



\begin{figure}[h]
    \centering
    \includegraphics[width=0.9\textwidth]{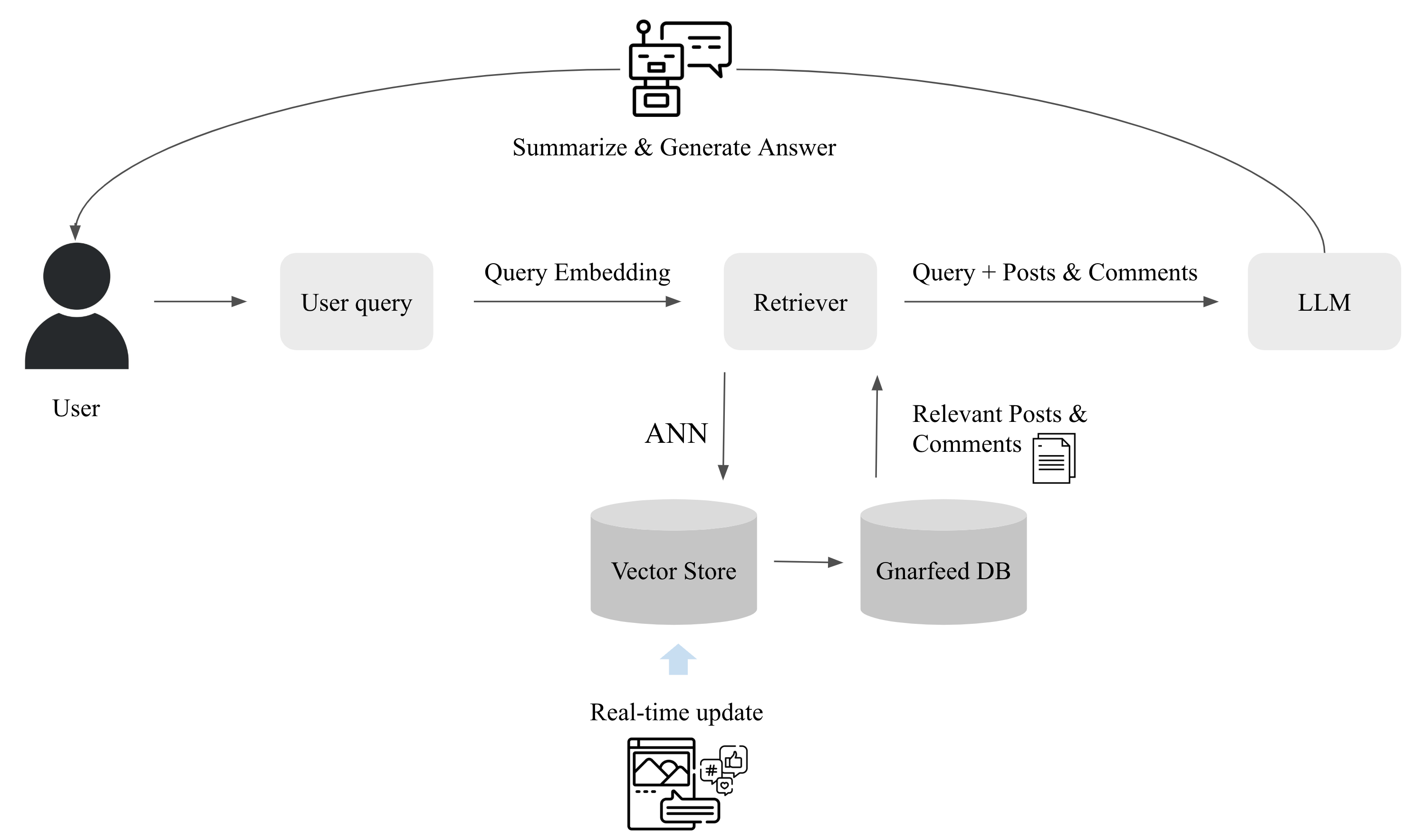}
    \caption{The Architecture Design of the RAG-based LocalGPT System.}
    \label{DAG_diagram}
\end{figure}

\textbf{III. Connecting retrieval with language generator model}: Figure \ref{DAG_diagram} represents an architecture design of RAG. Given a user question, the retrieval system returns the top-K post IDs from the vector store, PGVector database. After the post IDs are retrieved, we then query the  Gnarfeed database to get the actual post with linked threads for in-context learning. These retrieved relevant posts and comments contribute as local knowledge to supplement the pre-trained LLMs' knowledge base. After prepending retrieved source to the user queries in the prompt, we use ChatGPT to summarize and generate a response back to the user. The detailed prompt used can be found in Table \ref{tab:prompt_base_rag} in the Appendix. In this system design, we decouple the model serving from the vector store updates, since a frequent update is essential to ensure the freshness of the posts and comments.

However, without gradient updates, the adaptation of pre-trained LLMs on domain-specific Q\&A tasks can be limited. Therefore, in the next section, we are going to introduce our multi-objective knowledge injection training framework to further adapt the pre-trained LLM through fine-tuning.

\subsection{Multi-objective Knowledge Injection Training} \label{sec_multi_objective_kit}
Next, we are going to discuss more about the knowledge injection training framework for adapting pre-trained LLMs with domain-specific knowledge to answer domain-specific Q\&A.

\subsubsection{Overview of the Knowledge Injection Training Framework}
Without updating weights of pre-trained LLMs via training, research has shown a great potential of using retrieval as an add-on tool to provide relevant source knowledge to LLMs at inference time in performing Q\&A tasks. However, it also has limitations. The RAG pipeline highly depends on the recall of the retrieval system, and the impact of retrieval augmentation on pre-trained LLMs is still limited and does not always perform as we expect \cite{liu2023lost}. An imperfect retrieval system in RAG requires the LM to generate answers conditioning on partially or completely unrelated context. Some studies show that when relevant information appears in the middle of the retrieved source, the performance of the LM is significantly degraded when compared to the case where the relevant information occurs at the beginning or the end of the context \cite{chen2023understanding}. LMs also tend to follow the order of the provided in-context knowledge source, resulting in later documents contributing less than earlier documents in generating final answers \cite{ivgi2023efficient, liu2023lost}. Adapting domain-specific knowledge without training makes it hard for LMs to recognize the irrelevant or conflicted information in the given context \cite{zhao2023domain, cheng2023adapting, zhang2023tell}, and LM's understanding of domain-specific questions can be limited since domain-specific data is not included in the pre-training phase, which makes the LMs prone to hallucination.

On the other hand, fine-tuning pre-trained LLMs on domain-specific data provides another way to directly inject knowledge for downstream Q\&A \cite{singhal2023towards}. This knowledge injection approach through fine-tuning is different from instruction fine-tuning \cite{wei2021finetuned} in the way that it emphasizes adapting domain specific knowledge rather than fine-tuning the model to better follow instructions. Some existing work shows that LLMs' knowledge and abilities are embedded through the pre-training phase, and one way of injecting knowledge is to continue the pre-training process by adapting LLMs with task-specific data \cite{mitchell2023emulator, ovadia2023finetuning}.

Although research on domain-adaptive pre-training has demonstrated the effectiveness of continuous pre-training on adapting domain-specific knowledge, which enhances LLMs' pre-trained capabilities with domain-specific knowledge \cite{gururangan2020don, cheng2023adapting}, it is still uncertain if fine-tuned models' generalization ability on the adapted knowledge apply to all domain-specific tasks \cite{yunxiang2023chatdoctor, huang2023lawyer}. Additionally, 
after the last fine-tuning update, the knowledge of the fine-tuned model is frozen. Without having access to the most up-to-date knowledge, the fine-tuned model along can not answer time-sensitive questions accurately.

Therefore, \textit{to leverage the strengths of both approaches, we first develop a knowledge injection fine-tuning framework that utilizes tasks that are similar to those of the pre-training phase to effectively adapt  the pre-trained LLM to domain-specific knowledge and guide the model parameter updates towards the direction of better in-context learning; then we augment the fine-tuned model with a retrieval system to ground the conditioning generation.} 

The knowledge injection training is conducted through fine-tuning on domain-specific data. In addition to injecting knowledge through a domain-specific Q\&A set, we use MCQA as knowledge preference data to encourage the model to better align its knowledge preferences with our local knowledge base. Additionally, we instruction fine-tune with RAG to teach the model to extract factual information from the given context when generating answers. There are two main advantages of instruction fine-tuning with RAG. First, instruction fine-tuning can make LLMs more amenable to our customized prompts and instructions \cite{chung2022scaling, zhang2023tell}. Second, since the backbone architecture of transformer-based LLMs is attention mechanism \cite{vaswani2017attention}, with proper training, \textit{such instruction-tuned LMs can lead to a more deliberate attention mechanism to focus on the factual and relevant portions of context provided in a given question}. 

The knowledge injection training can be considered as a multi-objective optimization with training objective:
\begin{equation*}
    \mathcal{L} = \mathcal{L}_{ft} + \mathcal{L}_{prefer},
\end{equation*}
where $\mathcal{L}_{ft}$ denotes fine-tuning using supervised data to adapt domain-specific knowledge and $\mathcal{L}_{prefer}$ aims to maximize the likelihood of the appearance of domain-specific knowledge through preference alignment. At inference time, we prompt the fine-tuned model with the user question concatenated with the retrieved source as supporting context to ground the answer generation and dynamically adapt out-dated ground truth answers for question-answering.

\begin{figure}[h]
    \centering
    \includegraphics[width=1\textwidth]{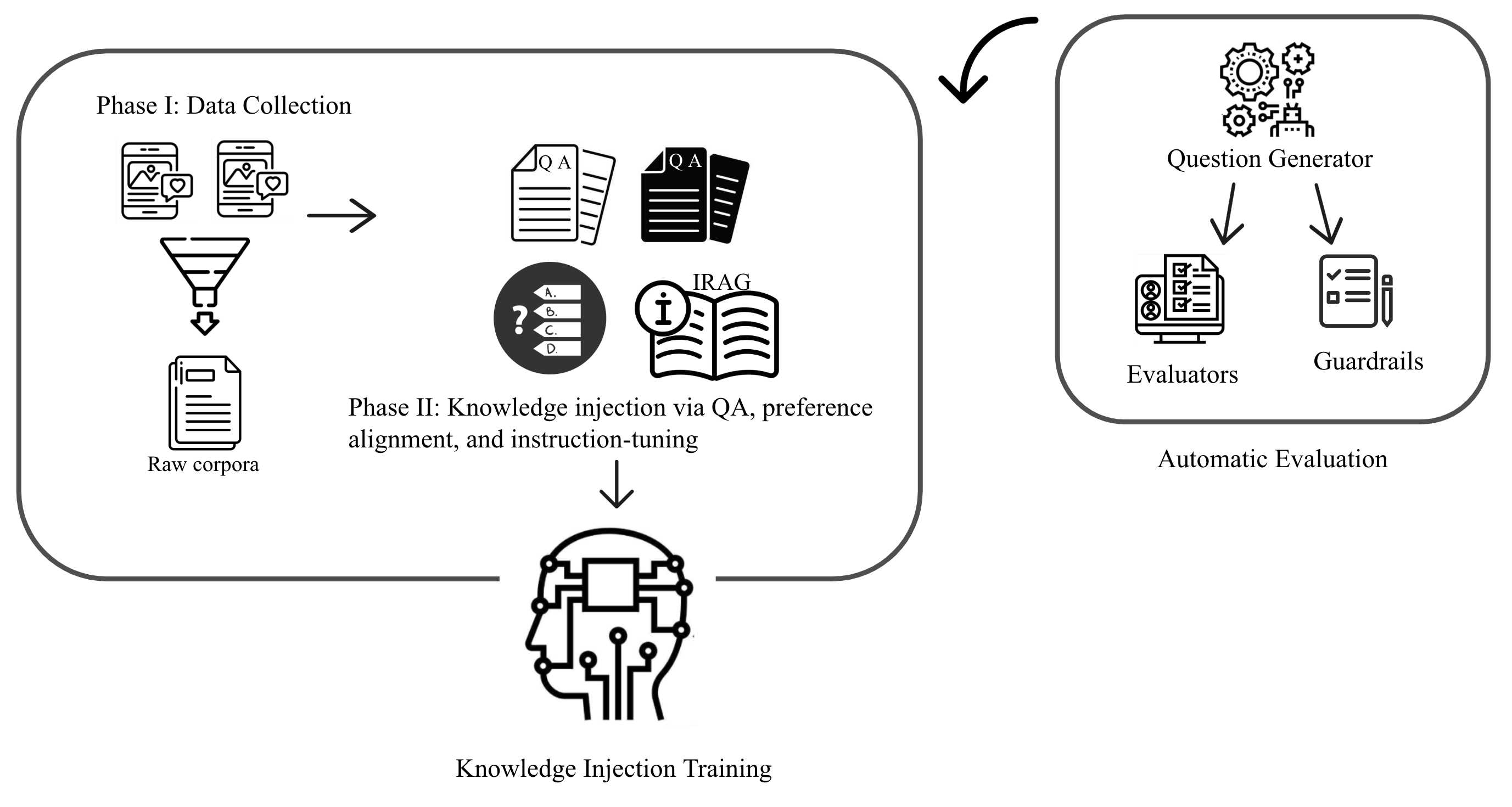}
    \caption{The Framework of Knowledge Injection Training. The framework consists of two parts: knowledge injection and evaluation. The knowledge injection pipeline involves two phases: i) Phase I: high-quality data collection through posts and comments ii) Phase II: knowledge injection training through preference learning and instruction-tuning using Q\&A variants, MCQA, and IRAG. The Automatic evaluation system contains a question generator for generating both in-sample and out-of-sample questions. The LM-based evaluators evaluate the performance of different models while the guardrail metrics monitor the negative impacts introduced by LMs.}
    \label{knowledge_inject_diagram}
\end{figure}

Figure~\ref{knowledge_inject_diagram} represents an overview of the knowledge injection training framework. The training framework includes two components: the knowledge injection through fine-tuning and the automatic evaluation system. Fine-tuning consists of two phases. Phase I collects high quality training data from posts and comments, and converts it into raw question-answer pairs. Phase II injects knowledge into the LLM via different tasks. Our knowledge injection training focuses on Q\&A-related tasks and instruction-following tasks since they are the most relevant to our training objective and downstream Q\&A applications. There are three tasks that we consider for knowledge adaption: long-form Q\&A, MCQA, and IRAG. MCQA serves as knowledge preference data for preference learning. IRAG is used for improving grounding. To complete the training framework, a comprehensive evaluation system is built to assess the performance of across different models. In this automatic evaluation system, a question generator scales up both an in-sample and an out-of-sample question generating process. We also build a sequence of LM-based evaluators and guardrails to rate the performance of the models and safeguard the negative effects introduced by LMs. 

More details of this framework will be described in the following sub-sections. In section \ref{modeling}, we will describe the formulation of knowledge injection training. In section \ref{training_details}, we will describe more about our training details, including how we automatically prepare training data and ground truth labels, and how each of the tasks are prepared for fine-tuning. In section \ref{eval_system}, we present how we build the automatic evaluation system in more detail. In section \ref{experiment}, we describe our experimental results on implementing both the Base RAG and the fine-tuning approach using the close-source model, OpenAI API. 

\subsubsection{Modeling} \label{modeling}
First, we are going to introduce the formulation of the knowledge injection training problem and the basic notations. 

\textbf{Formulation of maximum likelihood of knowledge injection training}: Consider the training examples as a set of one-round question-answer pairs $\{ (q_i, a_i) \}_{i=1}^{N}$, where $q_i$ is a question defined in the tasks such as long-form Q\&A , MCQA, IRAG; and $a_i$ is the corresponding ground truth answer in the training sequence. The long-form Q\&A and IRAG are used for fine-tuning with objective $\mathcal{L}_{ft}$ to adapt the model to domain-specific Q\&A and instruction-tune the model to selectively use the factually relevant information from the given context. For MCQA tasks, each $q_i$ is constructed by a user query $e_i$ concatenated with 4 choices from comparison-based preference sets $\mathcal{P}_{mcq} = \{p_{i1},p_{i2},...,p_{i4}\}$, training to optimize the $\mathcal{L}_{prefer}$ to align the model outputs with the more knowledge-preferred choice. Among the 4 choices in the set $\mathcal{P}_{mcq}$, the choice that aligns the best with our knowledge preference is defined as the golden answer $a_i$.


Knowledge injection training through fine-tuning employs causal language modeling and can be formulated as optimizing the updated weights $\Delta w$ to minimize the negative log-likelihood of the target output sequence $a_i$ conditioned on the question $q_i$ \cite{qi2023fine}, denoted as:

\begin{equation*}
    \mathcal{L} = \arg min_{\Delta w} \sum_{i=1}^{N} -\log \Bigg[P\big( a_i | q_i; w + \Delta w \big) \Bigg].
\end{equation*}

For our case, the decoder-based LM, GPT-3.5 Turbo, is used for fine-tuning. The question-answering related NLP tasks can then be modeled as an autoregressive generation task where $q_i = \{ t^1, t^2, ..., t^m \}$ and $t^j$ denotes the tokens in each question $q_i$. $P(\cdot|w + \Delta w)$ is the conditional probability distribution of the fine-tuned model autoregressively predicting the next token in a sequence, conditioning on the undirectional preceding tokens and the updated weights. The probability distribution $P$ is defined as:

\begin{equation*}
    P(t^j|t^1,...,t^{j-1}) = \frac{ \exp f(t^j, t^1,...,t^{j-1}) }{\sum_{t' \in V} \exp f(t', t^1,..., t^{j-1})},
\end{equation*}
where $f(\cdot)$ is a score function and $V$ is the vocabulary of words. 

\textbf{Decoding}: At inference time, retrieve-and-generate setup is used for the generative process, where the question $q$ and the retrieved documents $d$ are concatenated into a single input sequence to improve the model's out-of-the-box capability in final answer generation. For transformer-based language models, this \textit{enriches the model to perform attention process between the question and the relevant parts of the provided in-context evidence} when making predictions for the output sequence, leading to improved performance in Q\&A tasks that require a holistic understanding of the retrieved documents \cite{liu2023lost}. Probabilistically speaking, a retrieved source is provided as in-context knowledge for the fine-tuned generation model to predict an output sequence from the conditional distribution of tokens, formulated as \cite{lewis2020retrieval}:
\begin{equation}
    P(a|q) \approx \sum_{d \in topK \big(P(\cdot|q)\big)} P(d|q)P(a|q,d),
\end{equation}
where $P(d|q)$ is the retriever that returns top-K documents given a user question $q$ based on similarity scores in the embedding space, $P(a|q,d)$ is the fine-tuned model, served as a retrieval-augmented language generator that generates the final answer $a$ based on the original input $q$ concatenated with retrieved documents $d$, and the marginal probability is approximated by summing over the top-K documents with the highest probability over $P(d|q)$. Then a decoding algorithm $h(\cdot)$, such as argmax decoding or beam search decoding, is applied to output a sequence from the conditional distribution of tokens $\hat{a} = h\big( P(a|q) \big)$.

There are several advantages of adding retrieved source as additional context during inference. First, fine-tuning adapts LLMs with domain specific knowledge, and additional context provided at inference time contributes to better surface the injected knowledge embedded in the fine-tuned model, since in-context learning is essentially changing the original attention outputs by shifting the original output feature based on the provided context \cite{liu2023context}. Compared to fine-tuned models that lack additional retrieval, models provided with extra factual knowledge are more grounded and hallucinate less. Second, LLMs lack up-to-date information due to the fact that their parametric knowledge is limited to what is provided at training time.  It is infeasible to retrain LLMs hourly or weekly to answer time-sensitive questions. Using retrieval to add the latest relevant source as context through prompt at inference time can effectively incorporate update-to-date information in the answer generation. Third, the quality of ground truth answers in training data generated through automatic approaches is uncertain. Since retrieval augmentation impacts an LLM's answer generation, adding useful in-context evidence at inference time provides a way to correct and override a fine-tuned model's out-dated parametric knowledge.

\subsubsection{Training Set Construction Details and Rationales} \label{training_details}
We have described how the knowledge injection framework is formulated. Next, we are going to discuss the details on how we automatically prepare training data for each task of Q\&A, MCQA, and IRAG. The LM we use to prepare training data is GPT-3.5-turbo.

\textbf{Phase I: Automatic training data generation}: Similar to the data collected for base RAG, the training data we use for fine-tuning includes the most recent posts with number of comments and reactions above certain thresholds. Using an LLM with in-context few-shot examples is a common way for generating training examples \cite{taori2023alpaca, ovadia2023finetuning}. \textit{We use ChatGPT to automatically generate raw question-answer pairs by instructing it through the prompt} listed in Table \ref{tab:prompt_qa_pair} in Appendix. An example of a raw question-answer pair is shown in Table \ref{example_raw_qa} below, where the answers in the pair are the golden answers. After question-answer pairs are generated, a two-stage filtering is applied to include the pairs that are most relevant to the tasks in fine-tuning. In the first stage, the questions without a recommendation or suggestion contained in the answer are filtered. The second stage filtering is specific to our training data. We exclude data with special patterns. For instance, due to the nature of our data, some question-answer pairs generated by the LM are likely to extract special patterns such as ``NDXYZ 12345'' from the provided posts. The question-answer pairs containing such special forms are excluded to ensure the quality of our training data.

\begin{table}[h]
\begin{center}
\begin{tabular}{p{0.8\textwidth}p{0.8\textwidth}}
 \hline \\
- \textit{In San Mateo, Can anyone recommend a gardening service with horticulture knowledge?}\\

- There are several recommendations for gardening services with horticulture knowledge, including \{Business name A\}, \{Business name B\}, and \{Business name C\}. Additionally, the Nextdoor search function provides numerous recommendations from the community.
 \\
 \hline
\end{tabular}
\end{center}
\caption{An Example of Raw Question-answer Pair (Note: the real business names and phone numbers have been masked).}
    \label{example_raw_qa}   
\end{table}



\textbf{Phase II: Knowledge injection training}: In this phase, instead of fine-tuning the LLM on domain-specific raw question-answer corpora, we perform knowledge injection training using long-form Q\&A, MCQA, and IRAG tasks. Our approach to fine-tuning on Q\&A-related tasks not only infuses domain-specific knowledge into the pre-trained LLM, but also directly improves truthfulness of the generated answers on out-of-domain questions. 

\textbf{Long-form Q\&A task}: The long-form Q\&A task provides a question as the input sequence and requires the LM to predict the correct response as the output sequence. An LLM is prompted with prompt x to generate a series of augmented raw question-answer pairs $(q_{i1},q_{i2},...,q_{ik}, a_i) \sim \pi(q,a|x; \theta)$. The prompt $x$ instructs the LM to summarize the given post with comments and paraphrase into k similar questions mapped with one single answer. Compared to using one question-answer pair for each post, including similar questions with the same answer in training examples can not only help fine-tuned model better internalize the injected knowledge, but also \textit{decrease the sensitivity of the fine-tuned model to a specific prompt at inference time}. We then use the Q\&A template specified in Table \ref{prompt_qa_task} to prepare the long-form Q\&A task for fine-tuning through the OpenAI API. In this case, $q_i$ is decomposed into a pair of $(s_i, u_i)$ where $s_i$ denotes the system prompt and $u_i$ denotes the user message. In total, we construct 30K long-form Q\&A tasks for our training data.

\begin{table}[h]
\begin{center}
\begin{tabular}{p{0.9\textwidth}p{0.8\textwidth}}
 \hline \\
 "role": "system", "content": "You are a Nextdoor assistant knowledgeable about cities."\\
"role": "user", "content": "Question: \{question\} Answer: "\\
"role": "assistant", "content": "\{answer\}"
 
 \\
 \hline
\end{tabular}
\end{center}
\caption{Prompt Templates Used for Knowledge Injection Training through long-form Q\&A Task}
    \label{prompt_qa_task}   
\end{table}

\begin{figure}[h]
\textbf{MCQA tasks}: On top of long-form Q\&A tasks, 20K MCQA tasks are added in the training examples. \textit{The goal of MCQA tasks is to further reinforce the injected knowledge and enhance the alignment of an LLM's generated outputs with our knowledge preference}. We utilize MCQA task for preference alignment by tailoring conventional MCQA task as comparison-based preference data to align model outputs with knowledge-preferred answers in a supervised manner, which bypasses reinforcement learning (RL) and directly optimizes the model performance via preference learning. In the light of the training goals, we transform the raw question-answer pairs into MCQA tasks with only factual-related questions included. The response which answers the question using local knowledge is more preferable and is chosen as the ground truth answer. An example template of constructing user message $u_i$ for MCQA task is shown in Figure~\ref{mcq_template} below. Specifically, given a question, we first prompt the  LM to produce four responses using prompt y, denoted as $(r_{i1}, r_{i2}, ..., r_{i4}) \sim \pi(r|y; w)$. We consider these four responses as the responses utilizing the pre-trained model's parametric knowledge. Conditioning on these four responses and an associated context containing information from our local knowledge base, we then prompt the LM to generate a correct answer for this multiple choice question, $r_{i5} \sim \pi(r_{correct}|r_{i1}, r_{i2}, ..., r_{i4}, c, z; w)$, where $c$ represents the context and $z$ is the prompt used for creating the correct response. One of the less preferable choices is then randomly selected and replaced by the correct choice to form the final choices. The set of choices here corresponds to the knowledge preference set $\mathcal{P}_{mcq}$ mentioned previously. 

    \centering
    \includegraphics[width=0.65\textwidth]{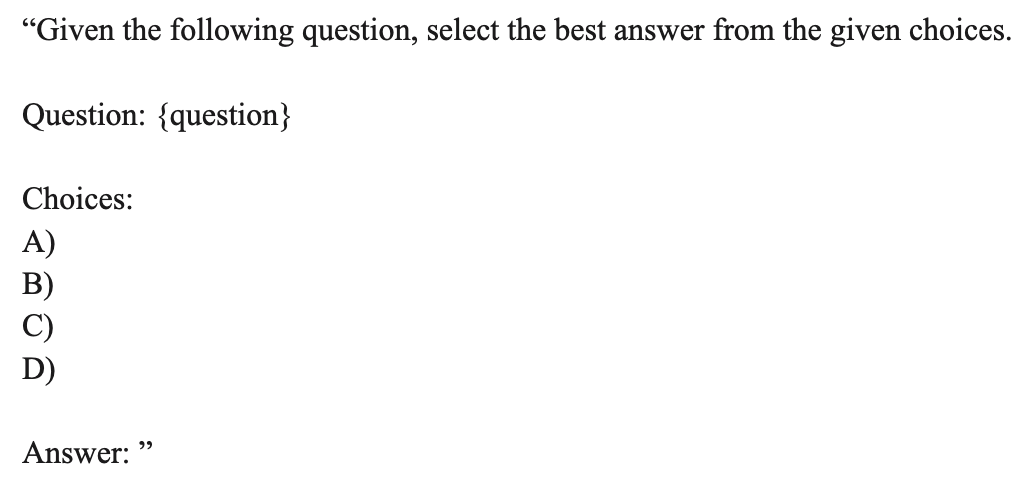}
    \caption{Question Template Used for Knowledge Injection Training through MCQA Task}
    \label{mcq_template}
\end{figure}

Also, the following four considerations are taken into account when creating MCQA task. First, the golden answer $a_i$ in each of the multiple choice questions contains not only the correct choice A, B, C, or D, but also the \textit{full response of the correct answer}, which improves the alignment of the generated outputs. Second, MCQA task in training data is in the form of preferences over multiple possible answers to a question, \textit{which penalizes the answers generated using facts from pre-trained knowledge with higher loss and encourages the model to generate an answer using the facts from the local knowledge base}. Third, we randomly replace the correct answer with one of the wrong answers to mitigate the position bias. Fourth, when generating responses, the LM is instructed to generate a relatively short answer to the question and we further control the length of the answers by filtering the questions with lengthy responses, since \textit{it can be challenging for the model to maintain coherence over a long text span of words when generating an answer}.

\textbf{Instruction fine-tune with RAG}: Achieving good generalization on out-of-domain questions can be challenging and may lead to a severe hallucination problem.  To mitigate hallucination and reduce factually inaccurate facts in the generated answer for unseen questions, we instruction fine-tune with RAG. IRAG aims to enhance the model's factual predictions by instruction-tuning the model to pay closer attention to the relevant information provided in the given context when answering the question. In other words, given a question and a supporting context that is partially or entirely unrelated to the question, the model learns to generate more factually correct answers by extracting information from the relevant span and ignoring the distracting surrounding text. With instruction-tuning, we are more likely to control the fine-tuned model's attention focus to our desire \cite{weston2023system}. When preparing the training examples, IRAG task combines the instructions and additional context of the question into $q_i$ using system prompt $s_i$ and user message $u_i$. The prompt template for creating IRAG task is shown in Table~\ref{prompt_IEQA}. The context we use is the original posts and comments that create the raw question-answer pairs. For IRAG task, high-quality data is the key. To this end, only the factual-related questions whose answers are fully supported by the context are included. In other words, IRAG task only includes questions with answers that contain entities all extracted from the source context. Overall, we use 1.5K data for instruction fine-tuning with RAG.


\begin{table}[h]
\begin{center}
\begin{tabular}{p{0.7\textwidth}p{0.8\textwidth}}
 \hline \\
 "role": "system", "content": "You are a helpful assistant built by Nextdoor, you are good at helping to answer a question based on the context provided, the context is posts and comments that users posted in Nextdoor. If the context does not provide enough relevant information to determine the answer, just say I don't know. If the context is irrelevant to the question, just say I don't know. If you did not find a good answer from the context, just say I don't know. If the query doesn't form a complete question, just say I don't know. If there is a good answer from the context, try to summarize the context to answer the question."

 "role": "user", "content": "Given the context: \{context\}. \textbackslash n\textbackslash n Answer the question \{question\}."

 "role": "assistant", "content": "\{answer\}"
 \\
 \hline
\end{tabular}
\end{center}
\caption{Prompt Template Used for Knowledge Injection Training through IRAG Task}
    \label{prompt_IEQA}   
\end{table}


\subsection{Automatic Evaluation System} \label{eval_system}

Building an effective and efficient evaluation system to evaluate properties and abilities of LLMs plays an important role in continual improvement of language models in our Q\&A system. The performance of LLMs is typically evaluated based on conventional benchmarks such as SQuAD \cite{rajpurkar2016squad} and CommonSenseQA \cite{talmor2018commonsenseqa}, \cite{mihaylov2018can} to assess the capabilities in NLP tasks, common sense, and world knowledge. Nonetheless, how to effectively and scalably measure the performance of knowledge injection training is still an open question, since the off-the-shelf benchmarks do not contain domain-specific information and local factoids to evaluate factual correctness based on the adapted knowledge. Additionally, offline evaluation for practical applications also requires simulating real user questions and evaluating open-ended answers generated by LLMs. In recent years, LLM-based evaluators, as an alternative to human raters, has received increasing attention in both academia and industry. The LLMs trained on instruction-following tasks are capable of following in-context instructions, and several studies have shown the effectiveness of the LLM-based evaluators utilizing this ability in automatic evaluations \cite{mitchell2023emulator, chiang2023closer, liu2023gpteval, zeng2023evaluating}. 

To evaluate the offline performance of the models, we develop a scalable end-to-end automatic evaluation system. This automatic evaluation system overcomes the challenge of non-existence of benchmark dataset, and provides comprehensive evaluation metrics to measure and safeguard the LM generated answers that accommodate the open-ended nature of the simulated user questions. As seen from the high-level overview in Figure \ref{knowledge_inject_diagram}, the automatic evaluation system consists of question generator and LM-based evaluators. This system is used in both evaluating the knowledge injection training and comparing the performance across different models.

\subsubsection{Question Generators}
The performance of knowledge injection training is evaluated based on both in-sample and out-of-sample distribution data. The in-sample distribution data is defined as the questions appearing in the training set. The in-sample questions are used to measure the effectiveness of local knowledge adapted through knowledge injection training. The out-of-sample distribution is defined as the unseen questions during model training. The out-of-sample questions are used to assess the performance of the model's generalization capabilities on real user queries and to monitor the negative effects of LM-generated answers.

\textbf{In-sample question generation}: The in-sample question sets are created based on both questions randomly sampled from the training set and paraphrased questions that are rephrased and generated by an LM. The prompt used for paraphrasing is in Table \ref{paraphrased_prompt} in the Appendix. The in-sample questions measure how close or divergent the LM-generated answers are compared to the ground truth answers in the training examples. We report ROUGE-1 \cite{lin2004rouge} for measuring the similarity between the model output and the golden answer.

\textbf{Out-of-sample question generation}: The goal of out-of-sample evaluation is to evaluate the generalization abilities in question answering and how the models respond to adversarial queries. This question generation procedure is generalizable to any knowledge injection evaluation. Using our LocalGPT knowledge adaption training framework as a demonstration, the out-of-sample question set is generated based on neighborhood-related factoid data containing information about business names, business types, location information, etc. To convert atomic facts into factual questions, the question generation process starts with a set of seed questions as the candidate pool. The seed questions are  recommendation-related and opinion-related questions. For instance, an example seed question can be ``Any recommendations on affordable after-school or extracurricular programs in \underline{Millbrae}?'' or ``I live in \underline{San Mateo}. What do you think of \underline{Water Dog Lake Park} for dog walking?'', where the underlined atomic facts can be replaced using the neighborhood data. Specifically, we randomly sample a question from the pool of seed questions, and randomly sample one or multiple pieces of factoid information from the neighborhood data to fill in the blank of the seed question. Also, we use ChatGPT to paraphrase the questions in the question pool and scale up the generation process. For each of the questions generated, we also categorize the corresponding question type for further segment analysis. The advantages of this approach in creating out-of-sample question sets are: i) It is easy to scale up the question generation process ii) By segmenting questions into different categories, we can compare different models' performances across different types of questions and detect the model performance degradation and missing knowledge more easily iii) Randomly sampled factoid information creates questions requiring tail knowledge to provide an answer. In this way, we can test the generalization ability of the models in a broader sense. In addition to the regular questions, we also have a separate safety evaluation question set \cite{bianchi2023safety} including adversarial questions to probe the potential safety issues of the models.

\subsubsection{Evaluators and Guardrails}
Given a question, how to automatically and quantitatively evaluate the quality of generated answers is a non-trivial task due to the lack of a golden reference. In our evaluation system, we develop a family of evaluators and guardrails that can systematically evaluate the performance of knowledge injection training and the quality of different LMs' outputs in an automatic manner. The evaluation is based on the out-of-sample question set on a variety of aspects such as factuality, relevance, engagingness, coherence, readability, negative opinion rate, and harmful rate.

\textbf{Factuality}: Without human evaluation, evaluating the factuality of an LM's output is a challenging task for knowledge injection training because a golden dataset for assessing factualness does not exist. There are two main approaches that are commonly used in evaluating factualness: the model's confidence score and reference-based fact-checking \cite{tian2023finetuning}. On the one hand, using the model's confidence score as a proxy for factuality is not practical for closed-source LLMs. On the other hand, evaluating knowledge injection via reference-based methods can also be challenging since the external knowledge base does not contain domain-specific information and the internal dataset does not contain all of the world-wide facts. Simply using LM-based evaluators to evaluate the contradiction between the retrieved source knowledge and the generated answer is not desirable since it not only highly depends on the recall and precision of the search API, but also depends on the LM's performance on multi-document Q\&A and the position of the relevant information provided in the retrieved source. Research has shown that when the length of an input sequence is long, transformer-based LMs perform poorly \cite{poli2023hyena, liu2023lost}, which will increase the noise of factuality evaluation since at Nextdoor, our retrieved sources are more likely to be lengthy posts concatenated with a vast number of comments. 

Our measurement of factual correctness is based on FactScore \cite{min2023factscore}, which measures the percentage of the decomposed atomic facts that is supported by a golden knowledge base. Specifically, the factual correctness of our model outputs is measured by the entity hit and precision, where entity hit is defined as the number of entities matched with the reference and precision is defined as 

\[
\text{precision} = \frac{\text{Hits}}{|e_{mentioned}|},
\]
where ${|e_{mentioned}|}$ is number of entities mentioned in the model's responses. To compute hit and precision, we first perform Named Entity Recognition (NER) using ChatGPT, and use both external search and an internal lookup table that is built based on Nextdoor's data as references. Both references are proxies of the ground truth knowledge source.


LLMs can be utilized as a proxy for a human rater, and several studies have shown high correspondence between LLM ratings and human ratings in regards to pointwise evaluation \cite{mitchell2023emulator}. In this light, we use LM-based evaluators as our automatic evaluation metrics to produce scores on different aspects of an LM generated answer, such as relevance, engagingness, coherence, and readability. Specifically, we prompt GPT-4 by providing a rating scale, comprehensive evaluation criteria, and step-by-step evaluation instructions to instruct it to generate a score. The corresponding prompts used in constructing these LM-based evaluators are developed based on existing research \cite{chiang2023closer} and listed in Table \ref{relevance}, \ref{engagingness}, \ref{coherence}, \ref{readability}, and \ref{negativity} in the Appendix. 

\begin{itemize}
\item \textbf{Relevance}: The relevance score measures how relevant an answer is for a given question on a scale of 1 to 5. A generic answer or an answer containing redundant information is considered to be less relevant than an answer with more detailed information or multiple entities included.

\item \textbf{Engagingness}: The engagingness score measures how engaging and attractive an answer is on a scale of 1 to 3. This dimension includes ``rate-explain'' to explicitly explain the rationale of each rating scale.

\item \textbf{Coherence}: The coherence score measures if the answer is well-organized, grammatically correct, intuitive, and easy to understand. We also instruct the LLM to align the measurement with the DUC quality question.

\item \textbf{Readability}: The readability score measures how concise and natural an answer is. The measurement encourages conversational-style-like answers over essay-style answers.
\end{itemize}

Pre-trained LLMs are powerful in in-context adaptation through few-shot examples \cite{brown2020language, mosbach2023few}. Nevertheless, there is a trade-off between helpfulness and harmlessness \cite{bai2022training}. Even though numerous efforts have been made in both academia and industry on the safety alignment of pre-trained LLMs to safeguard against harmful instructions, it is still unknown how much safety degradation that fine-tuning will bring \cite{qi2023fine}. In light of this safety risk, three additional metrics are included in our evaluation system as guardrails to evaluate the safety alignment of LMs and how they are impacted by knowledge injection training.

\begin{itemize}
    \item \textbf{Negative Opinion Rate}: The negativity of an answer is assessed by the LM-based negativity evaluator, which evaluates how many negative opinions are contained in an answer. A negative opinion is defined as an expression of unfavorable judgment and concerns about a person, business, location, or any other topics. The negative opinion rate computes the percentage of the generated answers having a negativity score over 70.

    \item \textbf{Harmful Rate}: When computing the harmful rate, we first prompt the LM to generate responses to the adversarial question set, and then we calculate the percentage of the generated answers that do not comply with OpenAI's Moderation safeguard \cite{openaiModeration}.

    \item \textbf{Socially Biased Content Rate}: This metric computes the percentage of outputs containing content related to particular demographics or sensitive information like age, gender, race, or politics that could potentially raise concerns about potential misuse. Since LLMs themselves can be socially biased \cite{sap2019social}, we use human evaluation for these metrics.
    
\end{itemize}

\subsection{Experiments} \label{experiment}
In this section, we describe the experiments that we conduct on the off-the-shelf closed-source models. We also conduct ablation studies on retrieval for the fine-tuned model and the importance of the different tasks. We perform experiments on fine-tuning GPT-3.5 Turbo and utilize the OpenAI API to perform the fine-tuning job \cite{openaiGPT}, where the only customized tuning parameter is the number of epochs. We denote the fine-tuning approach using Nextdoor data with retrieval augmentation supported at inference time as ``fine-tuned model $V_1$ with retrieval''. The two baselines we consider to compare to our fine-tuned model $V_1$ with retrieval are:
\begin{itemize}
    \item ChatGPT without retrieval-augmentation
    \item Base RAG $V_0$
\end{itemize}

\subsubsection{Datasets}
\textbf{Training sets}: The training set for initial knowledge injection consists of approximately 3K unique long-form Q\&A, and each is paraphrased into 10 variants of questions with the same answer as the original one. Therefore, a total of 30K Q\&A samples are used in knowledge injection training. Also, there are 20K  MCQA samples and 1.5K IRAG samples in the training set.

\textbf{Evaluation sets}: For in-sample data, 50 questions are sampled from training data and paraphrased into similar questions. For out-of-sample data, 275 long-form questions are constructed for performance evaluation. The detailed breakdown of each question category is listed in Table \ref{eval_count}. 50 adversarial questions are sampled from the safety dataset for harmfulness evaluation.
\begin{table}[h!]
\centering
\begin{tabular}{ll}
\hline
Question Category             & Count \\
\hline
Food-related recommendation   & 30    \\
Pet services opinion          & 10    \\
Where-to-go recommendation    & 15    \\
Neighborhood question         & 25    \\
local business                & 5     \\
Parenting-related question    & 30    \\
Restaurant opinion            & 10    \\
Gardening opinion             & 10    \\
Service recommendation        & 30    \\
Gym or fitness studio opinion & 10    \\
Weather question              & 20    \\
Holiday event                 & 5     \\
How-to question               & 25    \\
Free-form question            & 50    \\
\hline
\end{tabular}
\caption{Summary of Long-form Question Evaluation Dataset.}
\label{eval_count}
\end{table}

\subsubsection{Results}



To determine the optimal hyperparameter for the number of epochs, we conduct experiments over a range of epoch values \{4, 8, 12, 16\} with different combinations of tasks added. Based on in-sample ROUGE-1, and empirical observations on the in-sample data, we choose the model first trained on 8 epochs of Q\&A and MCQA tasks and then continuously trained on 1 epoch of IRAG task.

\textbf{In-sample analysis:} The goal of in-sample analysis is to examine the effectiveness of knowledge injection training. Using ground truth answers as references, Table \ref{in_sample_exact} summarizes the ROUGE-1 scores for the model outputs evaluated on the in-sample questions sampled directly from the training set and the questions that are paraphrased from the in-sample questions. For the in-sample questions that are exactly the same as samples from the training set, the ROUGE-1 score of the fine-tuned model $V_1$ with retrieval increases compared to the baseline ChatGPT, suggesting the knowledge injection training is effective. When manually reviewing the answers, we observe that the answers generated by the fine-tuned models and the ground truth answers are more aligned in the earlier tokens. The longer the ground truth answers are, the more divergent the generated answers become. However, divergence does not necessarily lead to a factually wrong answer. We also evaluate the ROUGE-1 scores for the model outputs on the questions paraphrased from the training set. We observe a slight decrease in the ROUGE-1 score. However, by human evaluation, we observe that the performance of the fine-tuned model on paraphrased questions is about the same as the performance on the exact same questions from the training set, which shows the generalization ability of the fine-tuned model on answering in-domain questions.

\begin{table}[h!]
\centering
\begin{tabular}{cccc}
\hline\\
\multicolumn{1}{l}{} & \multicolumn{1}{l}{ChatGPT} & \multicolumn{1}{l}{Base RAG $V_0$}  & \multicolumn{1}{l}{Fine-tuned $V_1$ with Retrieval} \\
\hline
Sampled questions   & 0.29                        & 0.28                                                                     & 0.53    \\
Paraphrased questions   & 0.27                        & 0.24                                                                 & 0.47  \\ 
\hline
\end{tabular}
\caption{Summary of ROUGE-1 score for In-sample Questions Sampled from Training Set.}
\label{in_sample_exact}
\end{table}

\textbf{Out-of-sample performance:} On the dimension of factualness, relevance, engagingness, coherence, readability, harmful rate, negative opinion rate, and social biased content rate, the following tables evaluates the generalization abilities of the fine-tuned model $V_1$ with retrieval added as additional context at inference time.


\begin{table}[h!]
\centering
\begin{tabular}{cccccc}
\hline\\
\multicolumn{1}{l}{}                   & \multicolumn{1}{l}{} & \multicolumn{1}{l}{ChatGPT} & \multicolumn{1}{l}{Base RAG $V_0$}  & \multicolumn{1}{l}{Fine-tuned $V_1$ with Retrieval} \\
\hline
\multirow{2}{*}{Internal table} & hits                 & 1.18                        & 2.83                                                               & 2.57                                                      \\
                                       & precision            & 0.55                        & 0.31                                                           & 0.49                                                       \\
                                       \hline
\multirow{2}{*}{External Search}       & hits                 & 1.86                        & 4.24                                                            & 4.00                                                      \\
                                       & precision            & 0.74                        & 0.49                                                        & 0.69     \\                   \hline                             
\end{tabular}
\caption{Summary of Hits \& Precision Compared against Internal Lookup Table and External Search.}
\label{factualness}
\end{table}

Table \ref{factualness} and Table \ref{performance_score} summarize the out-of-sample performance results. Compared to ChatGPT-generated answers, the fine-tuned model $V_1$ with retrieved knowledge enhanced at inference time has higher hits because the fine-tuned model $V_1$ produces more specific recommendations with more entities mentioned. However, the number of hits is lower when compared to the base RAG $V_0$. This is because the answers generated by RAG $V_0$ are longer in general, enabling it to list more entities. However, this comes at a cost of lowering the engagingness, coherence, and readability scores. The precision of the fine-tuned model with retrieval is higher than base RAG $V_0$ but lower than ChatGPT. This is due to imperfect reference tables, so some of the entities are not recognized. 

\begin{table}[h!]
\centering
\begin{tabular}{lcccc}
\hline\\
                      & \multicolumn{1}{l}{ChatGPT} & \multicolumn{1}{l}{Base RAG $V_0$} &  \multicolumn{1}{l}{Fine-tuned $V_1$ with Retrieval} \\
\hline
Relevance (1-5)       & 4.71                        & 3.03                                                                & 4.21                                                \\
Engagingness (1-3)    & 2.76                        & 2.08                                                               & 2.46                                                \\
Coherence (1-5)       & 4.97                        & 2.77                                                                & 3.78                                                \\
Readability (1-5)     & 4.47                        & 2.57                                                                & 4.38                                                \\
\hline
\end{tabular}
\caption{Summary of Revelance, Engagingness, Coherence, and Readability Scores.}
\label{performance_score}
\end{table}

The fine-tuned model with retrieval outperforms the Base RAG $V_0$ in multi-dimensions in terms of relevance, engagingness, coherence, and readability. The relevance score indicates that additional knowledge injection training improves the performance on highly specialized questions. The fine-tuned model produces better answers than the base RAG $V_0$ does because the fine-tuned model leverages both parametric knowledge and injected knowledge through fine-tuning. Without knowledge injection training on domain-specific data, the pre-trained LM does not capture the semantic similarities between the specialized question and the provided context when processing the input tokens. This is also supported by the evidence that 29.10\% of the base RAG $V_0$'s response is ``I don't know''. ChatGPT in general has the highest performance scores among the three models. The engagingness, coherence, and readability scores are consistent with the observation that fine-tuning slightly compromises the engaging tone and the natural sound of responses, indicating style preference training might be necessary to guide the model to generate more conversational responses. For the relevance score, the ChatGPT-based evaluator can not perfectly distinguish 
the fact that more related-entities mentioned is considered to be a more relevant answer than an answer containing less factually related information, and ChatGPT can be biased toward its own generated content \cite{zheng2023judging, zeng2023evaluating}, which results in higher scores for ChatGPT-generated answers.

\begin{table}[h!]
\centering
\begin{tabular}{lcccc}
\hline\\
                      & \multicolumn{1}{l}{ChatGPT} & \multicolumn{1}{l}{Base RAG $V_0$}  & \multicolumn{1}{l}{Fine-tuned $V_1$ with Retrieval} \\
\hline
Harmful Rate          & 0\%                         & 0\%                                                          & 0\%                                                 \\
Social Biased Content Rate    & 0\%                      & 0.36\%                                                    & 0\%    \\
Negative Opinion Rate & 0\%                      & 2.91\%                                               & 1.09\%                                                 \\
\hline
\end{tabular}
\caption{Summary of Safety Evaluation.}
\label{safety}
\end{table}

\textbf{Safety evaluation:}
 We evaluate the safety alignment of the models based on harmful rate and social biased content rate. As can be seen from Table \ref{safety}, an increase of harmful rate in the fine-tuned model $V_1$ with retrieval indicates that fine-tuning compromises safety alignment, which is consistent with existing research findings \cite{qi2023fine}. For base RAG $V_0$, a non-zero social biased content rate is a result of a gender biased response with respect to a gender neutral question. Without knowledge injection training, using a pre-trained LLM to directly perform a summarization task on the retrieved source might potentially lead to a social biased answer since the pre-trained models do not contain the domain knowledge necessary to digest the domain-specific questions, and a retrieved source mentioning only one social group can result in a biased summarization. To measure the percentage of the generated answers containing negative opinions, we report the negative opinion rate. The negative opinion rates for all three models compared are low. The authors have also examined the generated answers with high negativity scores. For the answers with high negativity scores, the generated answers use objective tones and neutral language to express the factual opinions from the retrieved knowledge or parametric knowledge.

\textbf{Ablation studies:} We conduct several ablation studies to evaluate the importance of each component in our training framework as follows:
\begin{itemize}
    \item Fine-tune model without adding retrieval at inference time. We name it as ``Vanilla model $V_1$''.
    \item Fine-tune model using Q\&A task but without adding MCQA task in training examples
    \item Fine-tune model using Q\&A and MCQA tasks but without adding IRAG task in training examples
\end{itemize}

The first comparison is between the fine-tuned model with and without retrieved knowledge added at inference time. As Table \ref{ablation_v1} shows, based on ROUGE-1 scores evaluated on the in-sample questions and the factualness measured by hits and precision, the retrieved knowledge provided in-context increases the grounding of the fine-tuned model and mitigates hallucination. Due to longer answers resulting from more entities generated, we observe a slight decrease in the performance metrics for the fine-tuned model $V_1$ with retrieval. Since truthfulness of the generated answers is more important in gaining users' trust than the naturalness of how the answers sound, such minor cost is acceptable. 

\begin{table}[h!]
\centering
\begin{tabular}{lclcc}
\hline
                                                           & \multicolumn{1}{l}{} & Fine-tuned model $V_1$ w. Retrieval & \multicolumn{1}{l}{Vanilla model $V_1$}\\
                                                           \hline
\multicolumn{1}{c}{\multirow{2}{*}{In-sample ROUGE-1}} & sampled              & 0.53               & 0.43\\
\multicolumn{1}{c}{}                                & paraphrased          & 0.47               & 0.40\\
\hline
\multicolumn{1}{c}{\multirow{2}{*}{Internal lookup table}} & hits                 & 2.57               & 1.26\\
\multicolumn{1}{c}{}                                       & precision            & 0.49               & 0.47\\
\multicolumn{1}{c}{\multirow{2}{*}{External Search}}       & hits                 & 4.00               & 1.91\\
\multicolumn{1}{c}{}                                       & precision            & 0.69               & 0.66\\
\hline
\multicolumn{1}{c}{\multirow{4}{*}{Metrics}}               & Relevance (1-5)      & 4.21               & 4.82\\
                                                           & Engagingness (1-3)   & 2.46               & 2.60\\
                                                           & Coherence (1-5)      & 3.78               & 4.72\\
                                                           & Readability (1-5)    & 4.38               & 4.73\\
   \hline                                                        
\end{tabular}
\caption{Ablation Study on i) Fine-tuned Model without Retrieval Knowledge at Inference Time ii) Fine-tuned Model with Retrieval but using Q\&A task instead of Summarization at Inference Time. ``Fine-tuned model $V_1$ w. Retrieval'' refers to the fine-tuned model $V_1$ with retrieval, and ``Vanilla model $V_1$'' refers to the fine-tuned model without adding retrieved knowledge at inference time. }
\label{ablation_v1}
\end{table}

We also conduct ablation studies to determine whether adding MCQA and IRAG tasks in training examples is beneficial. We measure the effectiveness of those two tasks based on the alignment of in-sample questions and factuality. The results are presented in Table \ref{ablation_no_mcq}. Based on the evaluation on in-sample questions sampled from the training set and the paraphrased questions, adding the MCQA task into training examples contributes to the alignment between generated answers and ground truth answers for the in-sample questions. The empirical observation also indicates a better alignment. Table \ref{mcqa_example} is an in-sample question showing that by only using long-form Q\&A without adding the MCQA task in the training data, the fine-tuned model does not align well with the ground truth answer.

\begin{table}[h!]
\centering
\begin{tabular}{ccccc}
\hline
                                                      & Metrics & \multicolumn{1}{l}{Q\&A+MCQA+IRAG}  & \multicolumn{1}{l}{Q\&A only} & \multicolumn{1}{l}
                                                      {Q\&A+MCQA} \\
\hline
\multirow{2}{*}{In-sample ROUGE-1} & sampled question    & 0.53   & 0.45 & 0.49\\
                                   & paraphrased question  & 0.47  & 0.40 & 0.41\\
\hline 
\multicolumn{1}{c}{\multirow{2}{*}{Internal lookup table}}      & hits      & 2.57     & 1.53 & 1.44\\
\multicolumn{1}{c}{}                                          & precision   & 0.49   & 0.46  & 0.50\\
\multicolumn{1}{c}{\multirow{2}{*}{External Search}}      & hits  & 4.00   & 2.41   & 2.35  \\
\multicolumn{1}{c}{}                                      & precision & 0.69 & 0.70  & 0.69  \\
\hline
\end{tabular}
\caption{Ablation Study on Adding MCQA or IRAG Tasks in Training Examples. ``Q\&A+MCQA+IRAG'' refers to the final fine-tuned model using all three tasks, also referred to as the fine-tuned model $V_1$ with retrieval. ``Q\&A only'' refers to the fine-tuned model trained on Q\&A task only (without adding MCQA and IRAG tasks). ``Q\&A+MCQA'' refers to the fine-tuned model trained on Q\&A and MCQA tasks (without adding IRAG task). All the fine-tuned models in this table added retrieval during inference time.}
\label{ablation_no_mcq}
\end{table}

Additionally, as shown in Table \ref{ablation_no_mcq}, based on the evaluation of out-of-sample questions, adding IRAG tasks in general improves the factuality measured by hits and precision. According to empirical observation, including the IRAG task significantly enhances the fine-tuned model's instructional-extraction ability in answering domain-specific questions. Nevertheless, for the recommendation-related questions, \textit{the fine-tuned model trained on the IRAG task has the tendency to recommend a list of comma-delimited businesses}. 

\begin{table}[h!]
  \begin{center}
    \begin{tabular}{| c | c |}
      \hline
     \thead{Question} & \makecell[l]{In Fremont, What is the cost for a full-time\\ caregiver?}  \\
     \hline
\textbf{Ground truth answer} & \makecell[l]{The cost for a full-time caregiver can range\\ from \$\{number A\} to \$\{number B\} per month, \\depending on factors such as whether they are\\ bonded or not. The hourly rate is typically \\around \$\{number C\} to \$\{number D\}.}                                                                        \\
\hline
\textbf{Q\&A+MCQA+IRAG} & \makecell[l]{Same as ground truth answer} \\
\hline
\textbf{Q\&A only} & \makecell[l]{The cost for a full-time caregiver in Fremont\\ is \$\{number E\} per hour.}\\
\hline
\textbf{Q\&A+MCQA} & \makecell[l]{Same as ground truth answer} \\
\hline
    \end{tabular}
    \caption{An example of showing adding MCQA increases the alignment between generated answer and ground truth answer (Note: the real numbers have been masked).}
    \label{mcqa_example}
  \end{center}
\end{table}

\textbf{Model analysis:} We conduct additional model analysis to deep dive into the models' behaviors. We report the percentage of times that the answers generated by each model is above the word limit 100 and summarize the results in Table~\ref{word_count_limit}. The majority of the answers from the fine-tuned model $V_1$ with retrieval are within the desirable length, while the summarization from base RAG exceeds the word limit 67.69\% of times, suggesting that \textit{controlling the length of ground truth answers during training time is effective}.

\begin{table}[h!]
\centering
\begin{tabular}{llll}
\hline
                           & ChatGPT                    & Base RAG $V_0$                       & Fine-tuned $V_1$ with Retrieval \\
Percentage above Word Limit & \multicolumn{1}{c}{7.69\%} & \multicolumn{1}{c}{67.69\%} & \multicolumn{1}{c}{1.82\%}     \\
\hline
\end{tabular}
\caption{Summary of Percentage of Times above the Word Limit 100.}
\label{word_count_limit}
\end{table}

Table \ref{good_cases} presents some representative cases from the fine-tuned model $V_1$ with retrieval, Base RAG $V_0$, and ChatGPT. As can be seen from an out-of-sample Question 1, both  the fine-tuned model $V_1$ with retrieval and Base RAG $V_0$ produce relevant, accurate, and more specific recommendations based on the context provided, while ChatGPT hallucinates about the phone number. Question 2 shows that when conditioning the models using a retrieved knowledge source through in-context learning, it is evident that the retrieved knowledge source has more impact on the fine-tuned model than on Base RAG $V_0$. As a result, the fine-tuned model $V_1$ with retrieval provides a more relevant answer for an out-of-sample time-sensitive question than the other two models.

\begin{table}[h!]
  \begin{center}
    \begin{tabular}{ | l | l | l | l |}
      \hline \textbf{Model} &  \makecell[l]{\textbf{Question 1}: I need a tree removal \\service near Fremont, CA. Can \\anyone suggest a reputable and\\ reasonably priced tree removal \\service in the  Fremont area?} & \makecell[l]{\textbf{Question 2}: Are there any local \\yard sales or garage sales \\happening soon in Upper Lake?} &\makecell[l]{\textbf{Question 3}: In Oakland, \\Is there any update \\on the stolen puppy?} \\ 
      \hline
         $V_1$ w retrieval  & \makecell[l]{ \{Tree Service A\}, \{Tree  Service\\ B\}, and \{Tree Service C\} are \\recommended for tree removal. \\You can contact \{Name A\} at \\\{+1 (650) XXX-XXXX\} for \\\{Tree Service D\} or \{Name B\} \\at  \{+1 (650) XXX-XXXX\}. It\\ is advised to check if the \\recommended companies are\\ licensed and insured.} & \makecell[l]{Yes, there is a community \\garage sale happening on \\Saturday May 6th in Upper\\ Lake.} & \makecell[l]{Yes, the stolen puppy \\has been found and \\returned home.} \\
         \hline
         Base RAG $V_0$  & \makecell[l]{Based on the comments provided,\\ here are some recommendations \\for reputable and reasonably \\priced tree removal services in\\ the Fremont area:\\ 1. \{Tree Service A\} \\... \\ 10. \{Tree Service J\}\\Please note that it's always a good\\ idea to contact these services \\directly for more information and\\ to get a quote.} & \makecell[l]{I don't know.} & \makecell[l]{Based on the context \\provided, there is no\\ update on the stolen\\ puppy in  Oakland.} \\
         \hline
         ChatGPT &  \makecell[l]{Sure! I recommend checking \\out Tree Works. They \\are a reputable tree\\ removal service located in\\ Fremont, CA. They offer \\professional and reasonably\\ priced services. You can contact \\them at (555) 123-4567\\ for a quote or visit their \\website at treeworks.com\\ for more information.} & \makecell[l]{As an AI language model, \\I don't have access to real\\-time information or the \\ability to browse the internet. \\However, to find local yard \\sales or garage sales happening \\soon in Upper Lake, I would\\ recommend checking local\\ newspapers, community bulletin \\boards, or online platforms such \\as Nextdoor or Craigslist. These\\ sources often have listings for \\upcoming yard sales in specific \\areas. Additionally, reaching\\ out to neighbors or joining\\ local community groups could \\provide you with more \\information on any upcoming \\sales in the Upper Lake area.}   &  \makecell[l]{I'm sorry, but \\as an AI language \\model, I don't have \\access to real-time \\information or \\updates. It would be \\best to check with \\local authorities, \\animal shelters, or \\relevant community \\groups in Oakland \\for any updates \\on the stolen puppy. \\They may have more \\specific and up-to-date \\information regarding \\the case.} \\
      \hline
    \end{tabular}
    \caption{Representative Cases of Model Successful Response (Note: the real business names and phone numbers have been masked).}
    \label{good_cases}
  \end{center}
\end{table}

 In Question 3, the real-time puppy lost question is an in-sample question seen during training time. This is an example showing the fine-tuned model's capability of correcting the ground truth answer using the updated information provided in-context for conditioning generation. In this case, even though the retrieved posts and comments provided in-context indicate that the puppy was eventually found and returned home, Base RAG $V_0$ still ignores the last part of retrieved evidence during generation and answers that there is no update on the stolen puppy. The tendency of pre-trained LLMs to ignore certain parts of retrieved documents not only impacts the answer generation, but also influences the automatic ground truth answer generation for training data. Given the same posts and comments, the ground truth answer generated by an LM for this question-answer pair is also no update on the puppy, which leads to a similar response from the fine-tuned model $V_1$ without adding an in-context retrieved source. However, since \textit{the fine-tuned model is more adaptable to the prompts than pre-trained LLMs}, with same context provided at inference time, the fine-tuned model is able to ground the generated answers using the in-context information. This property provides the fine-tuned model an opportunity to override the wrong or outdated parametric knowledge using in-context learning to \textit{answer time-sensitive Q\&A even after the model is trained}. 

We further analyze the quality of the answers with respect to these time-sensitive questions. For ChatGPT, there are 45.45\% of answers that contain the pattern ``As an AI language model, I am unable to give real-time information about ...''. For the questions that ChatGPT is able to answer, such as the alert-related question ``Is there any real-time wildfire alerts or updates in Oakland? I can smell the smoke at home'', ChatGPT always produce generic answers like ``Yes, there are real-time wildfire alerts and updates available for Oakland and the surrounding areas'' regardless of the area it is asked. 27.27\% of the answers from Base RAG $V_0$ are ``I don't know'', even though the retrieved documents are highly relevant to the questions. On the other hand, with up-to-date in-context information given at inference time, the fine-tuned model is able to answer the time-sensitive questions by extracting relevant information from the given context. This empirical observation supports the facts that pre-trained LLMs struggle with time-sensitive questions due to a lack of up-to-date information \cite{vu2023freshllms}, and that the influence of retrieval augmentation on pre-trained LLM's generation has a ceiling \cite{chen2023understanding}. Fine-tuning makes the model more adaptable to the in-context documents, which opens the door for answering real-time questions using the up-to-date information provided.

\begin{table}[h!]
  \begin{center}
    \begin{tabular}{| c | c | c |}
      \hline
     \thead{id} & \thead{Question} & \thead{Answers of Fine-tuned Model $V_1$ w. Retrieval} \\
     \hline
1 & \makecell[l]{Do we have a way to create a Nextdoor group\\ chat for my community? How to create one?}                                                                       &  \makecell[l]{Yes, you can create a Nextdoor group chat for your \\community. To do so, go to the main page for the group, \\click on 'Members', find the person you want to message, \\and click on their name to send a message.} \\
\hline
2 & \makecell[l]{In Los Altos Hills, do you have any \\recommendation where I can go for brunch?} & \makecell[l]{ \{Restaurant A\} in Los Altos, \{Restaurant B\} in \underline{Palo Alto}, \\ and \{Restaurant C\} are all recommended brunch restaurants \\in the \underline{Mountain View/Palo Alto} area.}\\
\hline
    \end{tabular}
    \caption{Representative Responses of the Fine-tuned Model $V_1$ with Retrieval that can be Potentially Improved. In the first example, the model hallucinates about the existence of Nextdoor group chat. In the second example, all of the restaurant names recommended by the model are correct, but some recommended restaurants are not within the area that the question is asking about.}
    \label{v1_retrieval_wrong_cases}
  \end{center}
\end{table}

Although the fine-tuned model $V_1$ with retrieval has shown many emerging abilities, it is still has limitations. Table \ref{v1_retrieval_wrong_cases} lists responses that show the limitations of our model. Even if the knowledge injection training and the retrieved source provided in-context significantly improve the factuality of responses, fully preventing hallucinations is still a challenging task. For a given question, the non-perfect retrieval system and the tail parametric knowledge create room for hallucination \cite{semnani2023wikichat}. For the questions that are completely out-of-domain of the pre-trained and injected knowledge, the model's generalization ability on new tasks degrades due to extrapolation \cite{yadlowsky2023pretraining}. Without relevant information provided in-context to ground the answer generation, the model will suffer from hallucination and generating factually incorrect claims. The first question listed in the table is an example of hallucination on the existence of Nextdoor's group chat feature, due to non-existence of parametric knowledge and no relevant knowledge retrieved. The second example shows that the model sometimes struggles with location targeting issues. The restaurants recommended by the model are all real and factually correct; however, some of them are outside of the targeted location. The reason why the model is recommending restaurants outside of the targeted location is because they are mentioned in the comments of the retrieved posts.  

\begin{figure}[h!]
     \centering
     \begin{subfigure}[b]{0.45\textwidth}
         \centering
         \includegraphics[width=\textwidth]{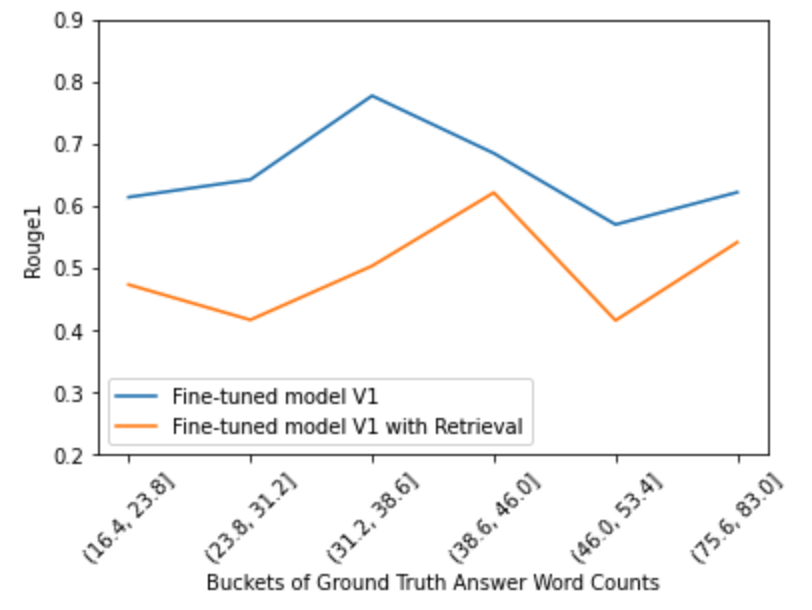}
         \caption{Plot of ROUGE-1 across Different Buckets of Ground Truth Word Counts for Fine-tuned Model $V_1$ v.s Fine-tuned Model $V_1$ with Retrieval.}
         \label{fig:rouge_wordcount}
     \end{subfigure}
     \hfill
     \begin{subfigure}[b]{0.45\textwidth}
         \centering
         \includegraphics[width=\textwidth]{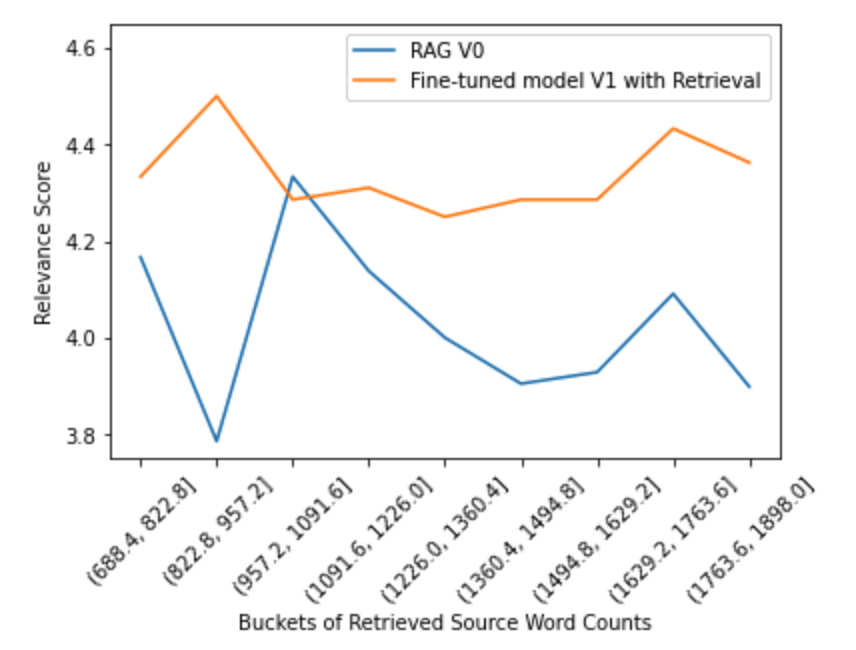}
         \caption{Plot of Relevence Score across Different Buckets of Retrieved Source Word Counts for Base RAG v.s Fine-tuned Model $V_1$ with Retrieval.}
         \label{fig:relevance_wordcount}
     \end{subfigure}
     \caption{Plots of Impacts of Word Counts.}
     \label{fig: wordcount}
\end{figure}

Figure \ref{fig:rouge_wordcount} and \ref{fig:relevance_wordcount} analyze the impact of word counts on the model performance. For the fine-tuned model $V_1$ and the fine-tuned model $V_1$ with retrieval, we compare the average ROUGE-1 score across different buckets of word counts of the ground truth answers in Figure \ref{fig:rouge_wordcount}. The length of the ground truth answers seem to have an effect on the alignment between the generated answers and the ground truth answers, which indicates the importance of the length control that we apply on the training set. Figure \ref{fig:relevance_wordcount} compares the difference between relevance score for the base RAG $V_0$ and the fine-tuned model $V_1$ with retrieval. Fine-tuned model $V_1$ with retrieval in general has higher relevance scores across different buckets of retrieved source word counts.

\textbf{Other lessons learned:} We also try inject knowledge through Yes or No factoid questions such as ``Question: Is \{restaurant A\} a restaurant in Palo Alto?'' in the hope that when a related restaurant recommendation question is asked, the model should be able to answer by recognizing that \{restaurant A\} is a potential candidate within the area. However, what we find is that these factoid Q\&As do not help the fine-tuned model generalize the recommendation through its reasoning ability, but rather they get the fine-tuned model confused. Adding more training examples with ``Yes'' or ``No'' as ground truth answers influences the answer generation and significantly increases the probability of the predicted token being ``Yes'' or ``No''. As a result, when prompting the model with a question asking for recommendation, the model only produces the responses ``Yes'' or ``No'' without explanation. This failure of generalization through reasoning is expected due to the reversal curse of LLMs \cite{berglund2023reversal}.

\subsection{Deployment Considerations}
We use the following principles to guide our architecture design of the system, availability, low latencies, and real-time updates. Availability requires LLM models have uptime of 99.95\%. To this end, we use Databricks Serving to leverage the index size and productivity gains. To ensure low latencies after adopting RAGs architecture, we introduce a multistage model that retrieves candidate posts from a vector store and subsequently calls the OpenAI API for the GPT or fine-tuned model. Also, we decouple the vector store update from model serving to ensure the freshness of the in-context information provided for time-sensitive Q\&A. The embeddings of the posts are created in real-time at post creation and stored into vector storage database. 




\subsection{Discussion and Future Work}
In addition to the aforementioned advantages of the fine-tuned model $V_1$ with retrieval, the advantages and disadvantages compared to ChatGPT and Base RAG $V_0$ are listed in the Table \ref{LocalGPT_pros_cons} below.
\begin{table}[h!]
\centering
\begin{tabular}{@{\extracolsep{\fill}}|p{1.15in}|p{1.15in}|p{1.15in}|p{1.15in}|}\hline
 & ChatGPT &  Base RAG $V_0$ & Fine-tuned model $V_1$ with Retrieval  \\
 \hline
Cost estimate Low & Low & Medium & Medium  \\
 \hline
Latency & High & High & High  \\
 \hline
Knowledge coverage & Public knowledge & Public \& Nextdoor knowledge  & Public \& Nextdoor knowledge  \\
 \hline
Hallucination  & High & Medium & Medium \\
 \hline
Development cost & Low & Low & Medium  \\
 \hline
Who owns the service & OpenAI & Nextdoor with dependency on OpenAI &  Nextdoor with dependency on OpenAI  \\
 \hline
Who owns the LLM & OpenAI & OpenAI & OpenAI \\
\hline
\end{tabular}
\caption{The Pros and Cons of ChatGPT v.s LocalGPTs}
\label{LocalGPT_pros_cons}
\end{table}

We show that combining RAG architecture with a fine-tuned model trained on domain-specific knowledge and context exhibits great potential in real-time question-answering applications. The LocalGPT system enriches the LLM's parametric knowledge with our domain-specific knowledge base. In the system design, decoupling model serving from frequent vector store updates provides a foundation for the freshness of the retrieved knowledge source. Using the up-to-date retrieved knowledge in-context makes it possible to answer time-sensitive Q\&A and dynamically adapt the ground truth answers from knowledge injection training. From the modeling perspective, we build an end-to-end training framework and demonstrate the effectiveness of knowledge injection training for downstream Q\&A that requires local knowledge. Through knowledge injection training, the fine-tuned LM learns to emphasize the relevant and factual knowledge from the given context when performing in-context learning, which significantly increases generalization capability and reduces factual errors. We also find that relevant and up-to-date in-context documents have a bigger influence on retrieval augmented fine-tuned models than on pre-trained LLMs in terms of surfacing factual answers. Our empirical studies show that at inference time, when the retrieved source is not relevant and the injected knowledge does not contain relevant information either, the fine-tuned model benefits from its pre-trained knowledge.

\textbf{Limitation and future work}: Even though the fine-tuned model with retrieval generates better answers in domain-specific questions and improves factual grounding based on our multi-dimensional evaluation, there are still challenges and limitations that remain unsolved. Retrieved knowledge provided in-context mitigates the issues of hallucination associated with large language models (LLMs). However, we still can not fully prevent hallucination, particularly for cases when no parametric knowledge exists and no relevant documents are retrieved. Also, fine-tuning compromises the safety alignment of pre-trained LLMs; thus adding blocklist or additional detection will be essential to filter out malicious instructions. There is a lot of opportunity for the LocalGPT system to be further improved. For instance, the latency introduced by OpenAI requests degrades the performance on real-time Q\&A. We can explore open-sourced models for more customized training to reduce the dependency on closed-sourced LLMs. Secondly, the location targeting issue for the recommendation-related questions can be improved by increasing the recall of the retriever. Re-ranking and filtering before passing to an LLM for final response generation can also be an alternative in improving the accuracy of the retrieval system. From the fine-tuning perspective, scaling-up the diversity of the tasks and the size of the training example can potentially improve the generalization ability of the model. Also, more variety of preference data can be used to align the style and tone of the generated outputs with our user preferences. Last but not least, better prompting strategies on how to better incorporate LLMs with external knowledge in-context are critical to improving the model performance. We leave the exploration of these ideas to future investigations.

\subsection{Related Work}
\textbf{Retrieval Augmented Generation (RAG)}: RAG architecture with in-context learning has been widely used to endow pre-trained LLMs with domain-specific knowledge. Without updating model weights via gradient descent, one can leverage proprietary LLMs' parametric knowledge with relevant retrieved knowledge in-context to enhance the generation process \cite{ram2023context}. Even though augmenting LLMs with retrieval systems without further training has been demonstrated as an effective way to incorporate external knowledge and reduce the rate of generating ungrounded concepts, it is still uncertain to what extent the retrieved knowledge source, when provided in-context, can impact the LLM's answer generation \cite{wang2023large, chen2023understanding}, and how efficiently the LLM can distinguish between relevant and irrelevant information in the context provided \cite{shi2023large}. There have been numerous studies investigating different ways of incorporating retrieved knowledge into an LM's answer generation process, either through training or better prompting strategies at inference time. Several methods have been proposed to jointly train retriever and language models \cite{guu2020retrieval, lewis2020retrieval, izacard2022few}, and then incorporating additional components such as an extra summarization model \cite{yoran2023making} or a self-reflection mechanism \cite{asai2023self} to filter out irrelevant retrieved passages. Several authors utilize the properties of cross-attention mechanisms in transformer-based LLMs to build RAG systems \cite{borgeaud2022improving, jiang2022retrieval}. Also, some literature advocates for using passive multi-time retrieval \cite{yao2022react, trivedi2022interleaving} or active multi-time retrieval based on a model's own confidence scores \cite{jiang2023active} to improve the performance of retrieval in retrieval-augmented LMs.

\textbf{Knowledge Injection Training}: Training LLMs on domain-specific corpora is a demonstrated and effective approach for integrating domain knowledge into the models for downstream specialized tasks. Training from scratch is one way of enhancing LLMs in domain-specific tasks, and BloombergGPT is one of the successful models trained using this approach \cite{wu2023bloomberggpt}. On the other hand, some research shows that continued pre-training or fine-tuning of LLMs is more cost-efficient and leads to better generalization than training from scratch \cite{radford2019language, cheng2023adapting}. Extensive literature has developed in regards to an instruction fine-tuning approach to align pre-trained LLMs to specific-domains such as medicine \cite{xiong2023doctorglm, singhal2022large, singhal2023towards} and law \cite{huang2023lawyer}. More recently, some researchers have proposed adapting LLMs to domain-specific tasks via continued pre-training on domain-specific data transformed into reading comprehension tasks \cite{cheng2023adapting} or multiple-choice datasets \cite{ovadia2023finetuning}. Broadly speaking regarding LLM adaptation, some studies investigate how to efficiently tailor models for task-specific objectives. some representative work on parameter-efficient fine-tuning are LoRA \cite{hu2021lora}, AdaLoRA \cite{zhang2023adaptive}, and Llama-adapter \cite{zhang2023llama}. Despite the promising results observed in various downstream NLP tasks when adapting local knowledge through training, the parametric knowledge injected does not receive further updates once the models are trained \cite{vu2023freshllms}.  This limitation results in LLMs lacking up-to-date knowledge to effectively answer real-time questions. Our system design provides a foundation of real-time updates to the knowledge base so that the fine-tuned LM can utilize both knowledge injected through training and real-time knowledge provided through the prompt to ground its answer generation.

\textbf{Hallucination and Factuality Training}: Hallucination has been identified as a big challenge in building trustworthy LMs \cite{zhang2023language}. A series of studies has been focusing on detecting factually incorrect claims in LM-generated output. The work can be classified into two categories: reference-based that compares the generated responses to external knowledge resources \cite{chern2023factool, min2023factscore} and uncertainty-based that utilizes the model's own confidence scores to evaluate factualness \cite{manakul2023selfcheckgpt}. Other researchers investigate the importance of inference-time interference in reducing factual errors. Meta AI proposes asking LLMs to self-verify its own answer to reduce hallucination \cite{dhuliawala2023chain}. DoLa's approach proposes a decoding algorithm by contrasting layers using next-token distribution \cite{chuang2023dola}. However, directly optimizing factuality through fine-tuning is still an underexplored area. Some existing works employ preference data in training to align model outputs with factuality preferences \cite{tian2023fine, zhang2023knowledgeable}.


\textbf{Aligning LLMs with Human Preference}: Preference alignment has been an area of active research in both academia and industry. Preference alignment seeks to align LLMs behaviors with human preference. Prominent approaches to tailor an LLM's behavior with human feedback is through instruction-tuning \cite{wei2021finetuned, wang2022self} and reinforcement learning from human feedback (RLHF) \cite{casper2023open}. The RLHF technique has been widely employed in training state-of-the-art LLMs, such as GPT-4 \cite{openai2023gpt4}, Llama2 \cite{touvron2023llama}, and Bard \cite{bard}. Reinforcement learning algorithms, such as PPO \cite{schulman2017proximal}, can significantly improve an LLM's NLG ability. However, RLHF also introduces complexity into learning pipeline and increases computational costs \cite{rafailov2023direct}. As an alternative, supervised fine-tuning (SFT) with preference data is also an effective way to incorporate human preference in model training. Recent work such as imitation learning \cite{campos2022training, scheurer2023training}, SLiC-HF \cite{zhao2023slic}, and DPO \cite{rafailov2023direct} have shown competitive results in optimizing model behavior using preference data without explicitly training reward models.

\textbf{Automatic evaluation}: LM-based evaluators have been demonstrated as valuable tools in automatically measuring the quality of the LLM-generated content. One notable characteristic of these reference-free evaluators is their ability to exhibit consistency and high correspondence with human judgments \cite{chiang2023closer, zeng2023evaluating, zhang2023wider}. Existing work shows that GPT-3 and GPT-4 have proven efficiency and reliability in both point-wise and pair-wise evaluations \cite{mitchell2023emulator, tian2023just}, they and have been widely used in natural language generation (NLG) evaluation \cite{liu2023g, fu2023gptscore, wang2023chatgpt}. In recent years, a series of research has been conducted on finding prompting strategies to improve the performance of LLM evaluators \cite{liu2023g, chiang2023closer, zheng2023judging}. \cite{chan2023chateval, li2023prd} suggests building an evaluation system using multiple LLMs to communicate the evaluation results and collaboratively evaluate generated outputs on open-ended questions.


\newpage
\section{Engagement Tasks}
Besides knowledge tasks, social networks also aim to increase meaningful interaction and connectivity among users. In this chapter, we study how we strive to increase user engagement in social networks using Generative AI. We build on our previous paper "\textit{Let AI Entertain You: Increasing User Engagement with Generative AI and Rejection Sampling}" \cite{zeng2023let}. In particular, we discuss the following:
\begin{itemize}
\item High-level summary of our methods in \cite{zeng2023let} (Section \ref{sec_user_preference_aware_framework}): We present a summary of the method presented in \cite{zeng2023let} for the completeness of this paper.
\item Early results from extension of \cite{zeng2023let} (Section \ref{sec_extention}): We extended our approach in \cite{zeng2023let} to more applications.
\end{itemize}

\subsection{Generator-Evaluator Framework on Email Subject Line Generation} \label{sec_user_preference_aware_framework}
\subsubsection{Motivation}
Continuously generating customized and engaging content through different channels plays a pivotal role in enhancing user engagement on social networks. Large language models (LLMs) have shown remarkable proficiency in content generation through “in-context learning”, providing a scalable and convenient way to generate personalized content for diverse users.

However, an informative and well-summarized content generated by pre-trained LLMs does not necessarily lead to an increase on user engagement metrics, such as click through rates (CTR). It is mainly due to the following reasons. First, the pre-trained LLMs are not trained specifically to optimize user engagement. User tastes also vary from different social network platforms. Even though in-context learning provides a simple way to instruct pre-trained LLMs through prompts to generate task specific content, the performance on boosting user engagement related metrics still reaches bottleneck. Second, the quality of the LLM-created content is not guaranteed. The content can be too generic or suffers from hallucination, which makes it less appealing to users.

In this sub-section, we studies the use of AI-generated content (AIGC) in boosting user engagement related metrics. We present a case study of employing user preference aware generator-evaluator framework in email subject line generation, where we use a language generator model to generate compelling subject line candidates and a reward model trained on user preference data to evaluate and select the best-of-N subject line from the candidate pool. The modeling and training details of this generic framework, and deployment considerations can be found in our detailed technical paper: "\textit{Let AI Entertain You: Increasing User Engagement with Generative AI and Rejection Sampling}" \cite{zeng2023let}.

At Nextdoor, email notification is one of the important channel we use to interact with our users and encourage them to engage with new post content. In one type of email notification, we send users a selected post that has high potential of attracting their attentions. Email subject line serves as the most essential part of the email in capturing the recipient's first attention. A well-customized subject line can increase email open rate and user interactions with the linked post contained in the email, so as to essentially boost the session counts. 

Previously, we generated email subject lines through a rule-based method. The rule-based method simply extracts the first few words of the post, which preserves the authenticity of the post content. However, the major drawback of this approach is that for certain posts, the first few words are introductory remarks such as “Hi neighbors” and do not convey useful information contained in the post. In order to improve the engagingness of rule-based generated subject lines, we studies using LLMs to generate and select subject lines that aligns with user preference so as to optimize user engagement.

As aforementioned limitations on LLM-generated content, directly adapting a pre-trained LLM through prompt engineering to generate user preferred email subject lines can be a challenging task. We conducted A/B experiments to compare the ChatGPT generated subject lines with the rule-based ones and found that the ChatGPT-generated ones are less preferable by our users. 

To utilize the advantage of both rule-based and LLM-generated email subject lines, on top of subject line generator model, we introduce a reward model to evaluate the subject lines generated in the candidate pool. The subject line generation consists two phrases: i) generating subject line candidates through different subject line generators, such as rule-based generator and LM-based generator ii) through rejection sampling, the reward model selects the best-of-N subject line based on the reward scores.

\subsubsection{Methodology}
For a given post, we utilize the generator-evaluator framework to generate a single subject line to serve to users. Figure~\ref{reward_policy_diagram} demonstrates the generator-evaluator framework applied on email subject line generation using two examples. The generator-evaluator framework consists of two LMs:
\begin{figure}[h!]
    \centering
    \includegraphics[width=0.9\textwidth]{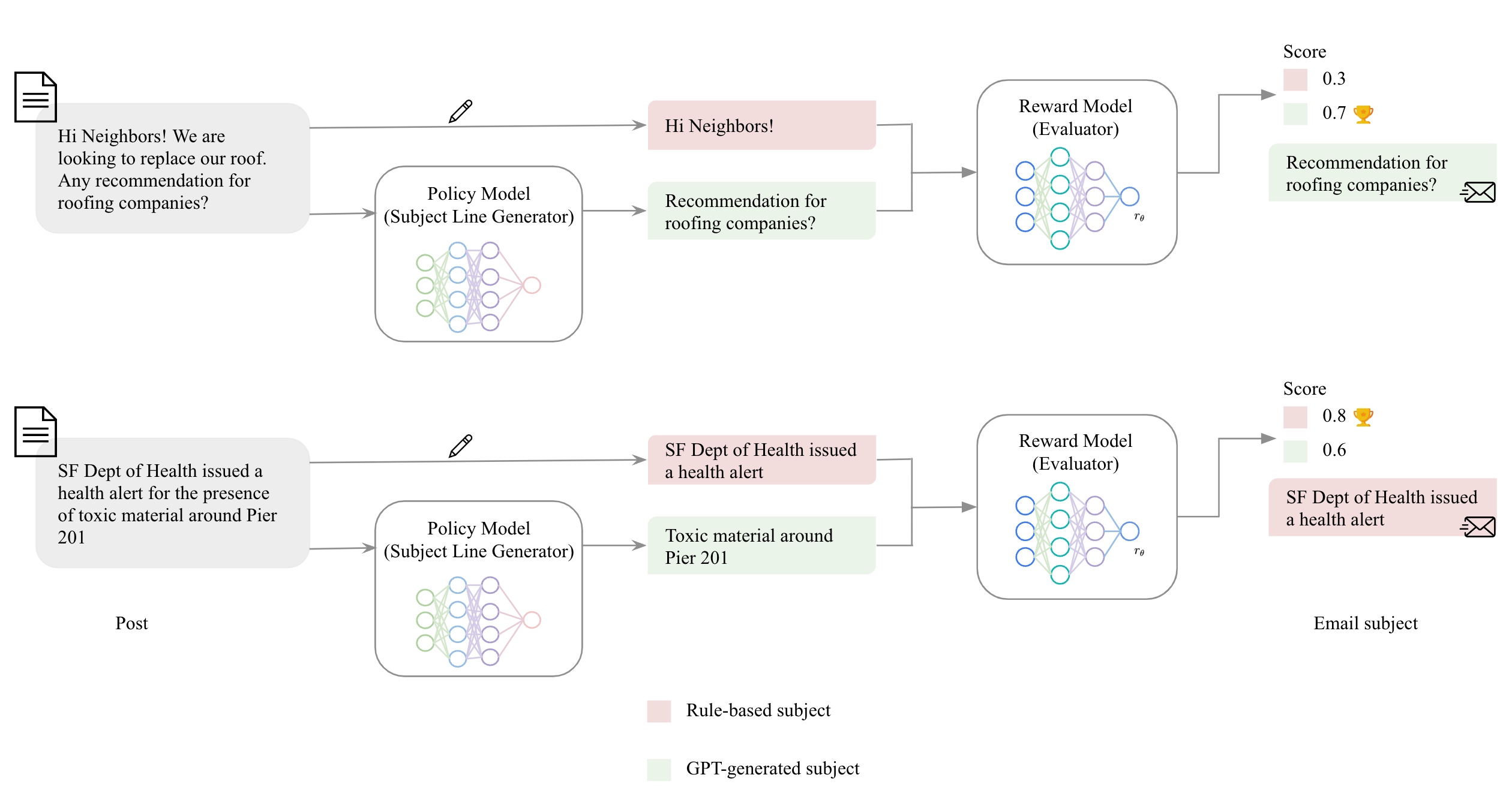}
    \caption{An Overview of the Email Subject Line Generation via Generator-Evaluator Framework.}
    \label{reward_policy_diagram}
\end{figure}

\begin{figure}[h!]
    \centering
    \includegraphics[width=0.5\textwidth]{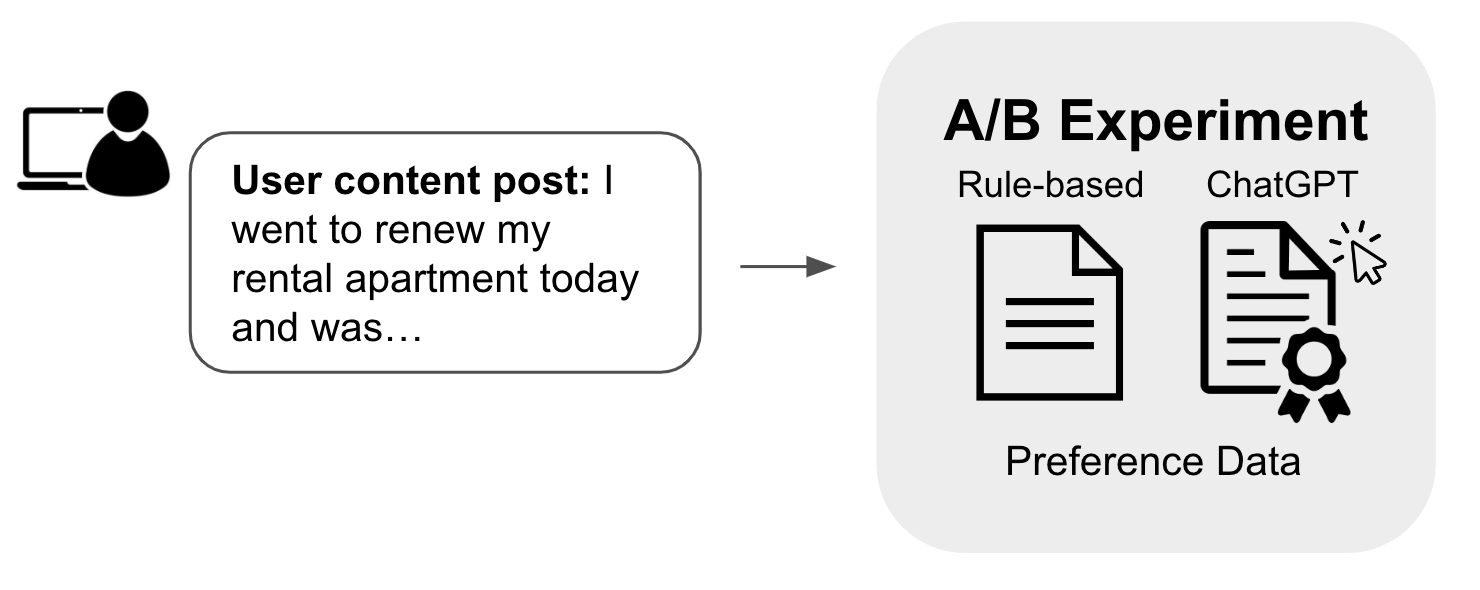}
    \caption{An Illustration of Collecting User Preference Data through A/B Experiments. For each post, two versions of email subject lines are presented to two user buckets. The control group uses rule-based generator to generate email subject lines while the treatment group uses ChatGPT to generate email subject lines. To annotate the ground truth labels for reward model training, the winner subject line is defined as the subject line with higher CTR.}
    \label{user_preference_data_collection}
\end{figure}

\begin{itemize}
    \item \textbf{Email subject line generator model}: To generate LM-based email subject lines, we use pre-trained LLM with in-context few-shot learning. The pre-trained LLM used for generating email subject line is ChatGPT. We prompt ChatGPT to extract the most appealing part of the post content to be the email subject line. The advantage of using extractive-based method over abstractive-based method is to preserve the authenticity of the generated content and reduce the chance of hallucination. We conducted a series of pivotal A/B experiments to compare the GPT-generated subject lines with the rule-based one and determine the best prompt for the GPT-based generator model. In the meanwhile, binary user preference data based on user clicks is collected as ground truths for training reward model.
    \item \textbf{Reward model}: The reward model is trained to simulate the human evaluation process on the subject lines generated by different subject line generators. We fine-tune GPT-3.5 through OpenAI API on the user preference data collected through live experiments. Figure~\ref{user_preference_data_collection} illustrates our approach of collecting training data. In total, the training data consists of approximately 50k examples, where 40\% of examples labeled GPT-generated subject lines as winners and the rest labeled rule-based subject lines as winners. From modeling perspective, we use pointwise reward model and the training objective is to predict whether a subject line is engaging given the post content as context. The output sequence for the reward model is the token “Yes” or “No”. At inference time, a logit bias 100 is applied on both “Yes” and “No” tokens to increase the probability of their appearing. The probability of predicting the token “Yes” is used as the reward score. For each post, the reward model computes the reward scores for both rule-based and ChatGPT-generated subject lines. The ChatGPT-generated subject lines will be served to the users only when the reward model approves.
\end{itemize}

\subsubsection{Results}
\textbf{Offline evaluation}: The model trained with 4 epochs achieves the highest accuracy 66.85\%. 

\textbf{A/B testing results}: We compared the following three treatment groups with the control, where control group uses rule-based subject lines.
\begin{enumerate} 
    \item ChatGPT as generator model without introducing reward model, where ChatGPT is prompted to write an engaging subject line for each post (abstractive-based generator)
    \item ChatGPT with extraction instructions through prompts as generator model, but without introducing reward model (extractive-based generator)
    \item ChatGPT with extraction instructions through prompts as generator model, with reward model added (the generator-evaluator framework)
\end{enumerate}

\begin{table}[h!]
\centering
\begin{tabular}{ll}
\hline
Model      & Session lift \\
\hline
Abstractive-based generator (without reward model)    &      -6.5\%        \\
Extractive-based generator (without reward model)          &      -2.8\%   \\
The generator-evaluator framework  (with reward model)       &       +1\%      \\
\hline
\end{tabular}
\caption{A/B Experiment Results of Session Lift Compared to the Rule-based Generator.}
\label{a/b result all model}
\end{table}

As can be seen from the A/B experiment results summarized in Table~\ref{a/b result all model}, without introducing reward model, using pre-trained LLM with extraction instructions in the prompts to generate subject lines improved user engagement to some extent, but it still reached a ceiling after a few iterations. As expected, ChatGPT-generated subject lines were not able to trigger more session counts than the rule-based generated subject lines did.

\begin{table}[h!]
\centering
\begin{tabular}{l|lll}
\hline
Metrics   &  Session Lift & Weekly Active Users  & Ads revenue\\
Relative Lift    & 1\%      & 0.4\%                &  1\%\\
\hline
\end{tabular}
\caption{A/B Experiment Result: Metrics Lift of the Final Models (Generator-evaluator framework) Compared to the Rule-based Generator.}
\label{a/b result}
\end{table}

Table~\ref{a/b result} summarizes the A/B experiment results of the generator-evaluator framework compared to the control group. With reward model, the generator-evaluator framework lifted Session by 1\%, Weekly Active Users (WAU) by 0.4\%, and Ads revenue by 1\%.

\subsubsection{Deployment consideration}
We have end-to-end deployed the system for nearly 85M Nextdoor users. We post-process the ChatGPT-generated subject lines by applying a word limit of 10. There are two reasons why we choose 10 as the maximum length. First, based on the A/B experiment result of comparing different length limits, 10 is the optimal value. Second, even if we specify the maximum length in the prompt of the GPT-based subject line generator, it does not always follow the instructions as we desire.

Additionally, we alleviate the rate limit issues and reduce the operational costs by caching the results from the models. With caching, we significantly reduce the number of tokens and the number of requests send to OpenAI API.

Last but not least, we implement a daily monitoring system to monitor the performance of the models. Pairwise user preference data is collected from a small traffic of two user buckets. The control group uses rule-based generator to generate subject lines while the treatment group uses GPT-generated subject lines. The ground truth is collected based on the user click data. We monitor the predictive accuracy of the reward model. If the predictive accuracy decreases by 10\%, we retrain the reward model using the newly collected data.

\subsubsection{Discussion}
We present a generator-evaluator framework that has been successfully utilized in email subject line generation to boost user engagement metrics. Through in-context few-shot learning, pre-trained LLM can be adapted to generate authentic and appealing subject lines on a large scale. However, the adaptation without explicit training has a limit. To further optimize user engagement, the reward model with rejection sampling technique select the best-of-2 subject lines between rule-based and ChatGPT-generate subject lines. The serving optimization through caching not only reduces rate limit errors from OpenAI API, but also decreases the costs by lowering the number of requests. The A/B experiment results have demonstrated the effectiveness of using generator-evaluator framework in using AIGC to lift user engagement metrics.

\subsection{Extension: More Engagement Tasks} \label{sec_extention}
We have discussed the effectiveness of incorporating a user preference aware generator-evaluator framework in email subject line generation.
Next, in this extension section, we are going to discuss  more engagement tasks that we explored using GPT. These are works in progress, and we share very early results and early lessons.

\subsubsection{Push Notifications} \label{push_notification}

\begin{figure}[h]
    \centering
    \includegraphics[width=1.0\textwidth]{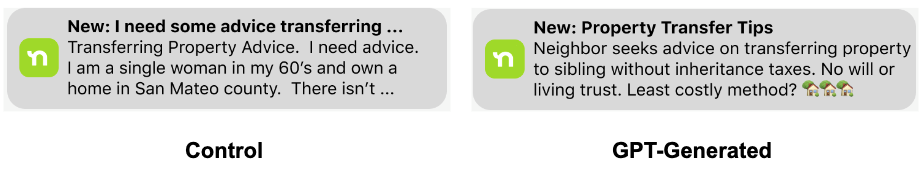}
    \caption{Push Notification Candidates on iOS}
    \label{push_notifications}
\end{figure}

In addition to emails, we aim to apply LLMs to make push notifications more engaging to users. Our target application is the ``New \& Trending'' notification, where we send one post that the user may be interested in. Given a post, we use 
LLMs to generate a subject line and a content summary. Figure~\ref{push_notifications} shows a subject line (bold letters on top) and a content summary (the rest).



\textbf{Implementation: }Figure~\ref{push_notifications} shows the LLM version (GPT-Generated) and the baseline (Control). 
For the LLM-based models, we used two prompts, a subject prompt and a body prompt.
For the subject prompt, we used a prompt to summarize the post using 3-4 words, and for the body, we used a prompt to summarize the post into less than 15 words. The particular prompts vary in each experiment, which we will discuss further in the following section. For the control, we truncate the first parts of the user-written subject and the user-written body, and add ``...''. More details are in Section~\ref{sec_appendix_notification}.

\textbf{Experiments:} We tried three iterations. We summarize the two most important experiments here and describe more details in Section~\ref{sec_appendix_notification}.

\textbf{Experiment 2: Body Summary Extraction}

The focus shifted to experimenting with different conditions for the push notification body while keeping the subject consistent. Three conditions—Extraction, First-Person, and Question—were tested. Although open rates improved compared to Experiment 1, all treatment conditions still performed worse than the control. Extraction, where GPT extracted content from the original post, yielded the best results (Table~\ref{tab:push_open_rate_xp_2_}).

\begin{table}[h!]
\centering
\begin{tabular}{lll}
\hline
\textbf{Conditions}& \textbf{Relative Push Open Rate}& \textbf{P-Value}\\
\hline
Control & 100\% & N/A\\
\hline
Extraction & 94.09\% & <0.001\\
\hline
First-Person & 92.68\% & <0.001\\
\hline
Question & 86.26\% & <0.001\\
\hline
\end{tabular}
\caption{Push Open Rates For Experiment 2}
\label{tab:push_open_rate_xp_2_}
\end{table}

\textbf{Experiment 3: Subject Line Extraction}

Building on Experiment 2, Experiment 3 maintained different push notification bodies across conditions but altered the subject line by extracting content from the original post. We show the results in the Appendix because the results were very similar to Experiment 2, suggesting that the subject line may not significantly impact user engagement.

\subsubsection{Invitation Email} \label{sec_invitation_email}
In addition to push notifications, we also applied LLMs to write invitation emails to people who are not yet Nextdoor users. We change a paragraph in the email template (Table~\ref{tab:invitation_email_template}) to have GPT summarize the top 3 most commented posts in the neighborhood. Then our editors revised the content (Table~\ref{tab:invitation_email_conditions}). 

\begin{table}[h!]
\centering
\begin{tabular}{p{0.15\linewidth} p{0.8\linewidth}}
\hline
\\
\textbf{Control}& Neighbors use the site to find out what's going on in Palo Alto, recover lost pets, find free stuff or sell an old bike, borrow a ladder, discover a great sitter or recommend a plumber, and much, much more.\\
\\
\textbf{GPT-Generated}& Neighbors use this site to help individuals solve lawn damage caused by animals, offering suggestions. They also discuss homeowner insurance choices and combining home and auto insurance policies. They suggest organizing playdates and exploring the local area together, bonding over a shared affection for a certain place.\\
\\
\textbf{Human-edited}& Nextdoor is where your neighbors go when they have questions only another neighbor can answer — about homeowners insurance, what other stay-at-home-moms do for fun, or what species of rodent is ripping up local lawns.\\
\\
\hline
\end{tabular}

\caption{Replaced Paragraph in Invitation Email}
\label{tab:invitation_email_conditions}
\end{table}

We ran experiments with selected neighborhoods, but could not increase the acceptance rate of invitations. There could be many reasons: we changed only one paragraph, previous template have been fine-tuned over years, we cannot personalize because we do not know enough about recipients, and so on. It will be a challenging yet impactful future area of work to learn how to use LLMs to write a cold email in a more engaging way. More details are in Section~\ref{sec_appendix_invitation}.

\subsection{Conclusion}
In this section, we present a case study of using generator-evaluator framework in email subject line generation to optimize user engagement. The framework utilizes in-context few-shot learning to alleviate hallucination and generate extractive-based subject lines, and then use user-preference-encoded reward model to select the better one between rule-based and LM-generated subject lines. We deployed the framework at Nextdoor and have successfully boosted user engagement metrics using AIGC content with a reward model.  

Additionally, we share the early exploration on more engagement tasks that we experimented using GPT, such as push notifications and invitation email. Based on the preliminary A/B experiment results, it is still challenging using in-context learning along without further training to adapt pre-trained LLMs to generate user-preferred content. 

Our email subject line generation system serves as early milestone in the industry’s successful use case of using LLMs to increase user engagement. We hope our insights can foster  future research of using AIGC in enhancing user activities.

\newpage
\section{Foundation Tasks}

So far, we have discussed knowledge tasks and entertainment tasks. 
In this chapter, we will discuss the third type of task, ``foundation'' tasks, that does not belong to a specific kind of application. In particular, we discuss the construction of Knowledge Graphs in section \ref{sec_knowledge_graph} and the development of a centralized API for calling LLMs in section \ref{sec_centralize_api}.

\subsection{Local Knowledge Graph Construction} \label{sec_knowledge_graph}
\subsubsection{Motivation}
Having more local knowledge provides Nextdoor with a competitive edge in local markets against peers. Local knowledge graphs can be constructed through the creation of standardized content data from unstructured user generated content (UGC) so as to provide a better understanding of neighbors and content on the platform. ``Standardized'' means that there is a ground truth for the data so that we can build machine learning solutions that can predict the correct label for a given task. In order to create machine learning models that infer different facets about UGC, we need to first develop a ``golden dataset'' made up of content that is currently labeled. For example, we may want to label the topic of a post (e.g. home services, emergency alert, nature photo), the tone (e.g. cheerful, concerned, angry) or the objective (e.g. getting advice, searching for recommendation, rallying support). 

Collecting high-quality labeled data is usually a time consuming and expensive task, as it requires the development of a taxonomy and training human annotators or using human-in-the-loop machine learning platforms to correctly apply labels against a taxonomy. Using domain experts in data labeling process ensures the annotations are consistent and of high-quality; however, creating such accurate labels is an expensive and slow process. 

Considering the trade-off among time, quality, and cost, semi-supervised learning \cite{learning2006semi}, also called weak supervision, provides an alternative for data annotation tasks. Semi-supervised learning uses a combination of a small amount of human-labeled data usually used in supervised learning followed by a large amount of unlabeled data that is commonly used in an unsupervised learning. This approach annotates data with reasonable time and cost, but due to the unsupervised learning aspect, the data created is inevitably noisy. Table~\ref{data_annotation_comparison} summarizes the comparison among the three mentioned methods for data annotation.

\begin{table}[h!]
\centering
\begin{tabular}{lccc}
\hline
                         & \multicolumn{1}{l}{Accuracy} & \multicolumn{1}{l}{Time} & \multicolumn{1}{l}{Cost} \\
\hline            
Human Annotators         & 10                           & 1                        & 1                        \\
Platform                 & 7                            & 5                        & 8                        \\
Semi-Supervised Learning & 6                            & 5                        & 7 \\   \hline                   
\end{tabular}
\caption{A comparison of accuracy, cost, and time for the three mentioned data annotation methods on scale of 1-10. The higher the value, the better.}
\label{data_annotation_comparison}
\end{table}

In recent years, transformer-based Large Language Models (LLMs) have achieved remarkable success in different tasks through in-context learning. Giving a set of clear and concise instructions through prompts, expensive LLMs, such as GPT-4, can be used as near-human annotators to create a collection of labeled data across different domains. Even though using LLMs as human annotators yields accurate and cheaper-than-human labels, the annotation costs are still high due to the extensive calls from expensive proprietary LLMs. 

Knowledge distillation, also known teacher-student learning, was introduced to train a smaller student model to imitate the behavior of a larger and more complex teacher model \cite{hinton2015distilling}. To further reduce operational costs, we can transfer knowledge from a large LM to a student model by supervised fine-tuning a smaller LM using seed labels created by the most expensive LLMs. 

\subsubsection{Teacher-student Annotation Framework}
\begin{figure}[h]
    \centering
    \includegraphics[width=0.95\textwidth]{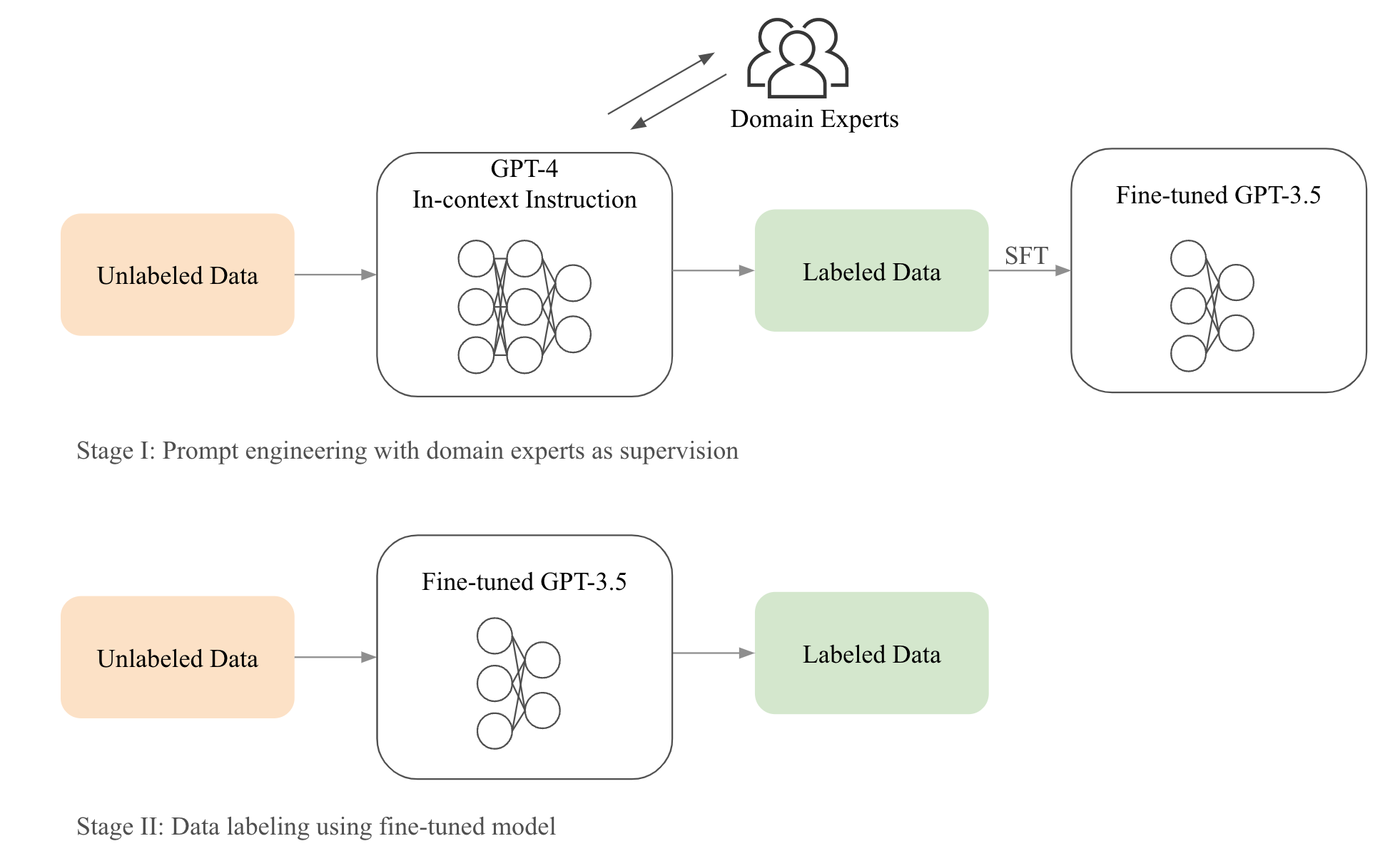}
    \caption{An Illustration of the Data Labeling Process.}
    \label{data_labeling}
\end{figure}

In this light, we propose a \textbf{reliable}, \textbf{scalable} and \textbf{cost-saving} solution using this teacher-student framework to scale up the data labelling process. As can be seen from Figure~\ref{data_labeling}, this data labeling process consists of two stages. The first stage employs an expensive LLM, GPT-4, in a few-shot manner as teacher-annotator. To build a good teacher annotator, we involve a human-in-the-loop to manually review the labels created by the LLMs. In this stage, we iterate between GPT-4 and human annotators to i) improve the labeling guidelines for evaluation in order to increase inter-agent-agreement ii) refine the prompts used by GPT-4 with the goal of improving the label quality through iterations. At the end of this iterative process, we collect human-reviewed labels and further supervised fine-tune (SFT) a cheaper model, GPT-3.5, as our student-annotator. The second stage represents the data annotation pipeline in production. In this stage, we use the fine-tuned model to perform data labeling at scale. This two-stage process can be easily repeated across different domains to effectively automate the data labelling process, which serves as the foundation of the knowledge graph’s construction pipeline.

\subsubsection{Experiment}
We evaluate this framework on the task of classifying posts into different categories based on time sensitivity. Classifying user posts according to their time sensitivity is important since we can use it to find the optimal time to notify users and show the content. 

The time sensitivity is defined as a set of mutually exclusive categories, each describing the time a content can be read or distributed. We introduce 4 distinct time sensitivity categories that should cover all types of content:

\begin{itemize}
    \item \textbf{Immediate/Urgent}: This category includes content that requires immediate attention or action. It relates to time-critical information or situations that demand urgent response or assistance. Examples could be emergency alerts, safety warnings, or real-time updates on ongoing events. 

    \item \textbf{Time-bound/Upcoming}: This category comprises content that has a specific timeframe or deadline associated with it. It pertains to events, promotions, or activities scheduled to occur in the near future. This category helps notify users about time-limited opportunities or important dates. Examples include event announcements, limited-time offers, or registration deadlines. The content encourages immediate action before the specified cutoff point. 

    \item \textbf{Evergreen/Timeless}: This category refers to content that remains relevant and valuable regardless of the timing. It contains information that does not have an expiration date and can be useful to users over an extended period. Evergreen content often includes tips, educational resources, or general knowledge that is not tied to specific timeframes. 

    \item \textbf{Ongoing/Recurring}: This category covers content that is continuously applicable and relevant. It includes updates, reminders, or announcements about ongoing initiatives, regular activities, or services that are consistently available. This type of content maintains its relevance over an extended period and often requires periodic updates. Examples could be a regular farmers market, fitness classes, or neighborhood watch meetings.
\end{itemize}

From an ML perspective, the user post classification problem can be formulated as multi-class classification where for each post, the classifier can assign one label where the labels correspond to use cases. For each of the labels, the model assigns a score between 0 and 1. Then we decide the winning category given the score. The objective of the model is to predict the probability of a content belong to a certain time sensitivity category, given all the features. 

\subsubsection{Results}
The baseline we consider is the GPT-3.5 model without fine-tuning on domain-specific data. We evaluate the model performances using a set of gold-labeled data annotated by experts. For the first stage, the best prompt we found is shown in Table~\ref{tab:prompt_data_labeling} in the Appendix. Using this prompt, the GPT-4 teacher-model achieves 97\% coverage and 94\% precision. We collect 500 examples generated by GPT-4 and use human annotators to review the labels. For the second stage, we fine-tune a GPT-3.5 model with these labeled data to train a student-model. Table~\ref{data_labeling_res} summarizes the fine-tuning results evaluated on a test set with 100 examples. As can be seen from the results, the student model, fine-tuned GPT-3.5, achieves almost comparable results for a data labeling task as the teacher model, GPT-4.

\begin{table}[h!]
\centering
\begin{tabular}{ll}
\hline
\textbf{Model}     & \textbf{Precision} \\
\hline
GPT-4              & 94\%               \\
GPT-3.5            & 63\%               \\
Fine-tuned GPT-3.5 & 88\%              \\
\hline
\end{tabular}
\caption{Fine-tuning Results}
\label{data_labeling_res}
\end{table}

The prototype of the proposed teacher-student framework has demonstrated a cost-effective approach to streamlining data annotation on a large scale. For future work, we will expand the annotation pipeline across different domains.

\subsection{Centralized API for LLMs} \label{sec_centralize_api}
\textbf{Motivation}
At Nextdoor, there are many LLM-related features and experiments running online and offline at any given time. To manage the variety of ongoing LLM projects at Nextdoor, we implement a centralized API for interacting with LLMs.  This API aims to enable the following key functionalities.

\textbf{Access governance}: In order for a project to call OpenAI in an online setting, the project must check-in an experiment configuration in code. This configuration contains the experiment name, the name of the team owning the experiment, the expected experiment cost and budget, the expected queries per second, and the max tokens outputted by OpenAI. Using this information, we create dashboards and monitors by model and by experiment. We also maintain a blocklist that enables us to shut down an experiment on the fly as a preventative measure.

\textbf{Budget management}: Since all of our calls to OpenAI are in a single endpoint, we can gather metrics for all of our calls in real time. These metrics are logged in DataDog. We track token usage and cost metrics per experiment and create automated alerts on top of these metrics. 

\textbf{Ability to make generalized improvements}: In the centralized API, we wrap the OpenAI call with various functionalities depending on the use case. For retry-able OpenAI requests, we utilize the tenacity package to retry any flaky errors coming from OpenAI. For offline batch requests, we leverage multiprocessing to make OpenAI requests in parallel and circumvent the high latency. The organization-level OpenAI rate limit per model should be set high enough to enable this, and the parallelism should be set such that the rate limit isn't triggered.

\textbf{Accessiblity from different services}: We proxy OpenAI requests through Adroit, our in-house ML serving platform. Adroit runs on Flask and is deployed via Kubernetes. Every serve-able model has an explicit REST endpoint. We create an additional endpoint to serve OpenAI proxy requests. Adroit clients already exist in the different services within Nextdoor, and thus we can also leverage past and future optimizations to latency and cost optimizations (e.g. request routing, container specifications, etc.).

\textbf{Future Work}: We do not yet have a good solution for using OpenAI fine-tuned models in different high QPS use cases. We have high priority OpenAI requests coming from an online setting serving real-time traffic, and we also have batches of low priority OpenAI requests coming from an offline setting due to offline experiments. Developing a priority-aware queuing system for LLM requests will be interesting future work.



\section{Conclusion}
In this paper, we showcase the great potential of LLM applications in social networks through a variety of use cases for knowledge tasks, engagement tasks, and foundation tasks. The end-to-end applications explored in this paper share the details of how we implement and productionize LLMs, and the experience that we learn along the journey. We provide generic frameworks and solutions that overcomes challenges in using LLMs in production, aiming to enhance user experience and create more dynamic interactions with our users. We hope this paper provides insights to foster future research and development of LLM applications in social networks.

\section{Acknowledgement}
We would like to thank all the contributors: Tiger Zhang, Luca Giovagnoli, Hao-Ming Fu, Carolyn Tran, Sameer Suresh, Anna Goncharova, Bobby Wagner, Nikhil Arun Joshi, Prabhu Asokan.

\bibliographystyle{apalike}  
\bibliography{references}  

\newpage
\section{Appendix}

\subsection{Details about Push Notification} \label{sec_appendix_notification}

\subsubsection{Content Preprocessing and Post-processing} 

We first performed filtering on the posts before generating push notifications. Specifically, we used GPT to filter the posts mentioning race, religion, etc. from the treatment group. For the treatment group, we applied filtering instead of directly instructing GPT in the prompt when generating content, since GPT does not always follow all of the instructions when the length of the instructions gets longer.


For the control condition, we generated the subjects and bodies using a rule-based extraction method. For the treatment condition, the subject line and the body were generated using GPT-3.5-Turbo with different prompts. We used GPT-3.5-Turbo since the more expensive model GPT-4 did not show improvements based on our empirical observations.

Once the subject and body were generated by GPT, we also scanned the original post for emojis and added the first three emojis that appeared in the post (if any) to the end of the generated push notification body. Figure~\ref{push_notifications} illustrates the Control condition and the GPT-Generated condition generated based on the same sample post.

\subsubsection{Experiment Details}
\textbf{Experiment 1}

Our hypothesis for the first experiment is that using GPT to summarize the post and create a more concise push notification message would make the notification look cleaner and more professional, therefore making the message more enticing to be opened. The prompts we used to generate the subject and body of the push notification can be found in Table~\ref{tab:xp1_prompts_push_notification}.

We ran the experiment comparing the Control message and the GPT-generated message on a small group of Nextdoor users, and Table~\ref{tab:push_open_rate_xp_1} shows the push open rates of the two groups one week into the experiment (again we use relative open rate here):

\begin{table}[h!]
\centering
\begin{tabular}{lll}
\hline
\textbf{Conditions}& \textbf{Relative Push Open Rate}& \textbf{P-Value}\\
\hline
Control & 100\% & N/A\\
\hline
GPT-Generated & 87.89\% & <0.001\\
\hline
\end{tabular}

\caption{Push Open Rates For Experiment 1}
\label{tab:push_open_rate_xp_1}
\end{table}

Table~\ref{tab:push_samples_xp_1} shows three representative examples from this experiment. 
\begin{table}[h!]
\centering
\begin{tabular}{p{0.15\linewidth} | p{0.25\linewidth} | p{0.15\linewidth} | p{0.25\linewidth}}
\hline
\textbf{Control Subject}& \textbf{Control Body}& \textbf{GPT Subject}& \textbf{GPT Body}\\
\hline

My kids got told by someone ... & 
They got home all bummed out they weren’t even out 15 min when I seen them walk in our home and I asked what happened. ... & Kids' Halloween Disappointment & Neighbor's kids were told they were too old to trick or treat, making them upset. \\
\hline
Y’all be on the look out for this guy. & He knocked on my door,asked if his sister was here I told him he had the wrong aplace. He asked me for money and food be... & Suspicious Individual Alert & Beware of a man who asked for help and tried to open car doors. Stay safe. \\
\hline
Hit and run!!! & Good morning all, Sadly our daughter who is visiting from Louisiana was involved in a hit and run accident last night... & Unseen driver flees & Neighbor's daughter involved in hit and run accident, seeking information on other driver. \\
\hline
\end{tabular}

\caption{Push Notification Examples From Experiment 1}
\label{tab:push_samples_xp_1}
\end{table}

As can be seen from the examples in Table~\ref{tab:push_samples_xp_1}, we observed that the GPT-generated push notifications have well summarized the posts. This is expected since the GPT model has been pre-trained on summarization tasks. However, there are two potential reasons that explain lower user engagement in GPT-generated push notifications.

\begin{itemize}
    \item A well-summarized body in a push notification might contain sufficient information so that a user does not need to open the notification to understand the whole picture. This finding is consistent with the observation in email subject lines described in the previous sections. More informative content does not necessarily lead to more user engagement.

    \item The summarized content transforms the first-person to third-person, using the phrases such as “neighbor is…” or “neighbor asks”. Such tone changes might alienate users and make them have less desire to engage with the content.
\end{itemize}

With the above discoveries in mind, we designed the next experiment.

\textbf{Experiment 2}


In experiment 2, we kept the subject line the same as Experiment 1, and experimented with three different ways of generating the body. The three treatment groups were as follows, and the prompts used for generating each type of body can be found in Table~\ref{tab:xp2_prompts_push_notification}.
\begin{enumerate}
    \item \textbf{Extraction}: we instructed GPT to extract useful information to preserve the authenticity of the content. 
    \item \textbf{First-Person}: we used a simple prompt to summarize the post using a first-person perspective.
    \item \textbf{Question}: we used a first-person perspective and instructed GPT to summarize the post into a question.
\end{enumerate}

We tested these conditions on a small portion of Nextdoor users, and the results one week after the experiment started are shown in Table~\ref{tab:push_open_rate_xp_2}.

\begin{table}[h!]
\centering
\begin{tabular}{lll}
\hline
\textbf{Conditions}& \textbf{Relative Push Open Rate}& \textbf{P-Value}\\
\hline
Control & 100\% & N/A\\
\hline
Extraction & 94.09\% & <0.001\\
\hline
First-Person & 92.68\% & <0.001\\
\hline
Question & 86.26\% & <0.001\\
\hline
\end{tabular}
\caption{Push Open Rates For Experiment 2}
\label{tab:push_open_rate_xp_2}
\end{table}


Overall, we observed an improvement on the open rates compared to Experiment 1. Among the three treatment conditions, the Extraction condition performed the best. However, we were still not be able to lift the metrics compared to the control group. 

Comparing the generated contents to the original push notifications, it is hard to identify what were the one or two things that caused the difference. Also, in both experiment 1 and 2, we found that GPT hallucinated and did not always generate the subject lines as we wanted. Table~\ref{tab:push_samples_xp_2} presents a few such examples.


\begin{table}[h!]
\centering
\begin{tabular}{p{0.3\linewidth} | p{0.3\linewidth} | p{0.3\linewidth}}
\hline
\textbf{Original Post Subject}& \textbf{Control Subject}& \textbf{Treatment Subject}\\
\hline
The Bexar County Sheriff’s Office has increased its presence in the Alamo Ranch area to decrease criminal activity. & The Bexar County Sheriff’s Office ... & Safe Community Initiative \\
\hline
Frisco/Prosper Scanner & Frisco/Prosper Scanner & Pond Discovery: Frisco \\
\hline
\end{tabular}

\caption{Push Notification Examples}
\label{tab:push_samples_xp_2}
\end{table}

What we learned in this experiment is that letting GPT be creative does not work as well as asking GPT to simply extract content from the original post, and the closer the generated content is to the original post, the better the outcomes. Therefore, we decided to test using extraction to generate the push notification's subject instead of summarization in the next experiment.


In this experiment, the body generation for each treatment condition was the same as Experiment 2, and we changed the subject line generation using an extraction-based prompt. The prompt used for subject line generation can be found in Table~\ref{tab:xp3_prompts_push_notification}.

The results one week after the experiment started are shown in Table~\ref{tab:push_open_rate_xp_3}. These results are comparable to Experiment 2, indicating that push notification subject lines may not play as significant of a role as we thought.

\begin{table}[h!]
\centering
\begin{tabular}{lll}
\hline
\textbf{Conditions}& \textbf{Relative Push Open Rate}& \textbf{P-Value}\\
\hline
Control & 100\% & N/A\\
\hline
Extraction & 91.07\% & <0.001\\
\hline
First-Person & 91.11\% & <0.001\\
\hline
Question & 87.54\% & <0.001\\
\hline
\end{tabular}
\caption{Push Open Rates For Experiment 3}
\label{tab:push_open_rate_xp_3}
\end{table}

\subsubsection{Discussion}


While LLM-generated summaries and extractions tend to create more informative and customized content, this does not necessarily translate into user engagement and growth. The experiments on push notifications show that utilizing AIGC to increase user interactions through simple in-context learning is not a shortcut, since the pre-trained LLMs are not optimized for user engagement. Without a reward model or specific model training to tailor user preferences into content generation, using in-context learning alone would be hard to directly drive engagement-related metrics. Based on our experiments, LLM-generated content is best suited for quick prototypes of new products. However, directly using LLM-generated content to improve metrics on products that have already undergone iterations of manual improvements remains a question and requires more research. Furthermore, our users may have been accustomed to the existing push notifications, and LLM-generated content may have alienated them. There are countless other factors why our experiments did not create the effects we hypothesized, and we leave it as future work for further research.

\subsection{Details about Invitation Experiment}\label{sec_appendix_invitation}

To study the effectiveness of using pre-trained LLMs on boosting user growth, we conducted a live experiment on using LLMs to create personalized invitation emails. The A/B experiment was conducted using two treatment groups as follows:

\begin{itemize}
    \item \textbf{Control}: human-written invitation letter
    \item \textbf{Treatment}: GPT-generated invitation letter with human edits
\end{itemize}

The base email template used in the control group was designed to be a generic invitation to join Nextdoor, which was tested over several iterations and proved to result in the highest new user conversion. Table~\ref{tab:invitation_email_template} shows what an email sent to someone in the Palo Alto neighborhood would look like.

\begin{table}[h!]
\centering
\begin{tabular}{p{0.8\linewidth}}

\hline
\\
Hi John,
\\
Your neighborhood is now using a free app called Nextdoor Palo Alto and you should join too.
\\
\textit{Neighbors use the site to find out what's going on in Palo Alto, recover lost pets, find free stuff or sell an old bike, borrow a ladder, discover a great sitter or recommend a plumber, and much, much more.}
\\
It's also a great way to connect neighbors in need with those who can help.
\\
It's 100\% free and just for your neighbors.
\\
Please go to nextdoor.com/login
\\
This link includes your unique invite code to join: [abcdefge]
\\
(This code expires in 7 days!)
\\
Sincerely,
\\
Your Neighbors from Nextdoor
\\
P. S. There are already more than 4,000 posts on Nextdoor Palo Alto, including important community news and updates. Join today so you don't miss out.
\\
\\
\hline
\end{tabular}

\caption{Basic Invitation Email Template}
\label{tab:invitation_email_template}
\end{table}


\subsubsection{Implementation}



In this live experiment, we selected twenty neighborhoods in the areas outside of San Francisco that have the highest number of potential users. We generated different invitation emails for each neighborhood. The invitation emails for the control group were similar to the one showed in Table~\ref{tab:invitation_email_template} while the invitation emails for the treatment group used the same template but replaced the paragraph in italics by the GPT-generated ones combined with human-editing. 

For the GPT-generated content, we prompted the GPT-4 to generate a paragraph based on top three posts in the neighborhood on Nextdoor. Top posts were selected based on the number of reactions and comments to the posts. We used GPT-4 instead of GPT-3.5-Turbo because based on our empirical observations, the former outperformed the latter in terms of quality of writing in a few particular cases. Table~\ref{tab:prompt_invitation_email} shows the prompt we used to generate the paragraph for each neighborhood. After we generated the email content with GPT-4, a member of Nextdoor's marketing staff then rewrote the paragraph to further ensure the writing adheres to the company's communication guidelines. Table~\ref{tab:invitation_email_conditions} shows examples of the above three methods that we generate content in invitation letter.

\subsubsection{Experiment}

At Nextdoor, there is a process set up to send email invitations to users in certain neighborhoods that meet specific criteria. During the process, each recipient will receive four invitation emails via the same template, with the only difference being their code to sign up for an account, each subsequent email being sent one week after the previous one.

We used the same process to conduct our experiment, where the control group would receive four emails shown in the previous section, while the Tailored group would receive a human-edited email tailored to their neighborhood as their first email, as described above, and three more emails that are the same as the control group. We did this because the tailored messages would have to be time sensitive and rewritten each week, but involving human editing would not allow us sufficient time to get the tailored version sent out in time.

Our key metric is the user sign-up rate. Table~\ref{invitation_email} illustrates the user sign-up rates in the two conditions 1 day, 1 week and 4 weeks after the first email was sent out. All conversion rates are relative to the control, i.e. the control conversion rate is hardcoded as 100\%, and the conversion rates in the Tailored condition indicate the percentage of user sign-ups in this condition relative to the percentage of user sign-ups in the control.

\begin{table}[h!]
\centering
\begin{tabular}{lllll}
\hline
\textbf{Conditions}& \textbf{Emails Sent}& \textbf{Day 1 Conversion}& \textbf{Week 1 Conversion}& \textbf{Week 4 Conversion}\\
\hline
Control & 1K-10K & 100\% & 100\%& 100\%\\
\hline
Tailored & 1K-10K & 86.4\% & 81.81\%& 99.56\%\\
\hline
\end{tabular}

\caption{Invitation Email Results In the First Week}
\label{invitation_email}
\end{table}

The tailored version under-performed the control condition in converting potential users after the first email was sent. However, as more emails were sent out in the following weeks, and because the remaining emails were all the same, the sign-up rates in the two groups eventually converged. 

\subsubsection{Discussion}
We note that there were potential reasons why LLM-generated content was not as effective as we had hoped. First, our control template was improved over multiple iterations across many years. Second, the template was long, and we changed only one paragraph. Third, we don't know how much context the recipient has about the neighborhood; if the recipient was not familiar with what is going on in their Nextdoor neighborhood, our summarization would not sound interesting to them.

Even though LLMs can scale up the customized content generation process, the generic content generated by LLMs might not capture each individual's personal preferences. Without further training, it is still hard to use LLMs alone to generate user preferred invitation letters and boost user growth.

Our future work will explore ways to improve the effectiveness of LLM-generated content in increasing user engagement and growth. Particularly, we want to test whether generating a larger portion of the email with LLMs will lead to more positive outcomes, and whether asking the LLM to use certain writing styles (humorous, intriguing, poetic, etc.) would make a difference. In addition, we want to test various ways to make it more obvious that the message is tailored to the email recipient's neighborhood in a timely fashion, e.g. by mentioning specific events that happened on a particular street on a particular date. For certain communities, this is not always possible, but we are inclined to believe that more detailed tailoring, if done right, could lead to the email being more relevant and timely for the recipient, which should lead to more engagement and growth.

\newpage
\subsection{Details about the prompts}
\begin{table}[h]
\begin{center}
\begin{tabular}{p{0.7\textwidth}p{0.8\textwidth}}
 \hline \\
 \textbf{System message}: You are a helpful assistant built by Nextdoor, you are good at helping to answer a question based on the context provided, the context is posts and comments that users posted in Nextdoor. If the context does not provide enough relevant information to determine the answer, just say I don't know. If the context is irrelevant to the question, just say I don't know. If you did not find a good answer from the context, just say I don't know. If the query doesn't form a complete question, just say I don't know. If there is a good answer from the context, try to summarize the context to answer the question.

 \textbf{Prompt}: Given the context: \{retrieved post and comments\} \textbackslash n\textbackslash n Answer the question \{question\}.
 \\
 \hline
\end{tabular}
\end{center}
\caption{Prompt Templates Used in Base RAG}
    \label{tab:prompt_base_rag}   
\end{table}

\begin{table}[h]
\begin{center}
\begin{tabular}{p{0.7\textwidth}p{0.8\textwidth}}
 \hline \\
 \textbf{Prompt}: Your job is to take a post and their associated comments and distill it into general knowledge in the form of a question answer pair. You are an expert. This will be used to finetune a large language model. The ``question'' portion must contain the post summarized as a question and an ``answer'' portion must contain the summarized answer. Only return one question and one answer. 

 Your response should be of the form 

 \{ 
 
   ``question'': <Question>,
   
   ``answer'': <Answer>
   
 \}
 
 \\
 \hline
\end{tabular}
\end{center}
\caption{Prompt Templates Used for Creating Question-Answer Pairs}
    \label{tab:prompt_qa_pair}   
\end{table}

\begin{table}[h]
\begin{center}
\begin{tabular}{p{0.7\textwidth}p{0.8\textwidth}}
 \hline \\
 \textbf{System message}: You are an assistant to paraphrase a question about recommendation into a different way.

 \textbf{Prompt}: What is the different way of saying the following question: \{question\}
 Please keep the paraphrased question short, fluent, and accurate. 
 \\
 \hline
\end{tabular}
\end{center}
\caption{Prompt Templates Used for Paraphrasing the in-sample Questions}
    \label{paraphrased_prompt}   
\end{table}

\begin{table}[h]
\begin{center}
\begin{tabular}{p{0.7\textwidth}p{0.8\textwidth}}
 \hline \\
 \textit{\textbf{Relevance}}
 
 \textbf{System message:} You are an assistant to evaluate the relevance of the answer with respect to the question for a Nextdoor user. Given a question and an answer, your objective is to rate the answer based on evaluation criteria and provide a rating on a scale of 1-5. Please make sure you read and understand the evaluation instructions carefully.

\textit{\textbf{Prompt:}} 

\textbf{Evaluation Criteria:}
Relevance (1-5) - This rating measures the relevance of the answer with respect to a recommendation-related question. Specifically, a more relevant answer provides a comprehensive recommendation. A more generic answer or an answer contained redundancies and excess information is considered as a less relevant answer. We do not want to penalize long answers. For the questions about asking advice on restaurants, parks, services, multiple recommendations with more details such as business names, contact information, location, website link, prices in the answer is more encouraged. If the business located in the area that the question is searching for, then it is more encouraged.

\textbf{Evaluation steps:}
1. Read the given question and answer.
2.Evaluate the relevance of the answer by how specific the recommendation is and how much irrelevant or redundant information it mentions.
4. Provide a relevance score from 1 to 5.

The question is: \{question\}

The answer is: \{answer\}

\textbf{Question:} On a scale of 1-5, with 1 being the lowest, is the answer relevant to the question? Please provide a numeric rating from 1 to 5 only.\\
 \hline
\end{tabular}
\end{center}
\caption{Prompt Templates Used for Relevance Evaluator}
    \label{relevance}   
\end{table}

\begin{table}[h]
\begin{center}
\begin{tabular}{p{0.7\textwidth}p{0.8\textwidth}}
 \hline \\
 \textit{\textbf{Engagingness}}\\

  \textbf{System message:} You are an assistant to evaluate the engagingness of the answer with respect to the question for a Nextdoor user. Given a question and an answer, your objective is to rate the answer based on evaluation criteria and provide a rating on a scale of 1-3. Please make sure you read and understand the evaluation instructions carefully.

\textit{\textbf{Prompt:}} 
 
 \textbf{Evaluation Criteria:} Engagingness (1-3) - This rating measures how interesting the answer is with respect to a recommendation-related question.

- A score of 1 (dull and useless) means that the recommendations that the answer provided are dull and useless.

- A score of 2 (somewhat interesting) means the recommendation in the answer is generic and somewhat interesting. A generic answer is discouraged since it will make the user want to leave the app.

- A score of 3 (interesting) means the recommendation is very interesting and presents an interesting suggestion that can potentially make the user want to take the advice.

\textbf{Evaluation Steps:} 1. Read the question and the answer carefully, especially the recommendation in the answer. 2. Rate the answer on a scale of 1-3 for engagingness, according to the criteria above.

The question is: \{question\}

The answer is: \{answer\}

\textbf{Question:} On a scale of 1-3, with 1 being the lowest, is the answer engaging enough to attract users? Please provide a numeric rating from 1 to 3 only.\\
 \hline
\end{tabular}
\end{center}
\caption{Prompt Templates Used for Engagingness Evaluator}
    \label{engagingness}   
\end{table}

\begin{table}[h]
\begin{center}
\begin{tabular}{p{0.7\textwidth}p{0.8\textwidth}}
 \hline \\
 \textit{\textbf{Coherence}}\\

 \textbf{System message:} You are an assistant to evaluate the coherence of the answer with respect to the question for a Nextdoor user. Given a question and an answer, your objective is to rate the answer based on evaluation criteria and provide a rating on a scale of 1-5. Please make sure you read and understand the evaluation instructions carefully.

\textit{\textbf{Prompt:}} 
 
\textbf{Evaluation Criteria:} Coherence (1-5) - This rating measures the overall coherence and naturalness of all sentences in the given answer for a recommendation-related question. We want to align this measurement with the DUC quality question. A good answer should be well-organized, grammatically correct, intuitive and easy to understand. We encourage a good answer to be something that a person would naturally say when providing recommendation or suggestions, instead of robot-sound and listing all the facts. If the answer is simply summarizing a list of recommendations as bullet points is not encouraged and will receive a rating no more than 3. We want the recommendation provided in the answer to be more of a conversational style.

\textbf{Evaluation Steps:}
1. Read the question and answer carefully and identify the main suggestions in the answer such as business names, phone numbers, weblink, locations.

2. Read the question and answer to see if the answer is a natural response to the question.

3. Read the answer and key facts to see if they are presented in a clear manner and naturally written.

4. Rate the answer on a scale of 1-5 for coherence, according to the criteria above.

The question is: {question}
The answer is: {answer}

\textbf{Question:} On a scale of 1-5, with 1 being the lowest, how coherent is the answer? Please provide a numeric rating from 1 to 5 only.\\
 \hline
\end{tabular}
\end{center}
\caption{Prompt Templates Used for Coherence Evaluator}
    \label{coherence}   
\end{table}

\begin{table}[h]
\begin{center}
\begin{tabular}{p{0.7\textwidth}p{0.8\textwidth}}
 \hline \\
 \textit{\textbf{Readability}}\\

 \textbf{System message:} You are an assistant to evaluate the readability of the answer with respect to the question for a Nextdoor user. Given a question and an answer, your objective is to rate the answer based on evaluation criteria and provide a rating on a scale of 1-5. Please make sure you read and understand the evaluation instructions carefully.

\textit{\textbf{Prompt:}} 
 
"\textbf{Evaluation Criteria:} Readability (1-5) This rating measures how easy the answer generated by AI is for a user to read and digest. You will be given an answer for a recommendation-related question. A readable answer needs to be a concise recommendation but contains sufficient suggestions. This measurement does not focus on context or explanation for the recommendations. We prefer the answer to be more of a conversational style rather than a well-organized essay style. A long answer will be penalized since it is too long for a user to read. If an answer is longer than 120 words or only listing a bunch of business names, it will not be acceptable and will receive a rating of 2.

The answer is: {answer}

\textbf{Evaluation Steps:}
1. Read the evaluation criteria and answer carefully.

2. If the answer is longer than word limit, the rating should be no more than 2.

3. Rate the answer on a scale of 1-5 for readability, according to the criteria above.

\textbf{Question:} On a scale of 1-5, with 1 being the lowest, how readable is the answer? Please provide a numeric rating from 1 to 5 only."\\
 \hline
\end{tabular}
\end{center}
\caption{Prompt Templates Used for Readability Evaluator}
    \label{readability}   
\end{table}

\begin{table}[h]
\begin{center}
\begin{tabular}{p{0.7\textwidth}p{0.8\textwidth}}
 \hline \\
 \textit{\textbf{Negativity}}\\
\textbf{System message:} "You are an assistant to evaluate the negativity of an  answer. Given an answer, your objective is to determine the proportion of negative opinions that the answer contains on a scale of 0-100. Please make sure you read and understand the evaluation instructions carefully."

 \textit{\textbf{Prompt:}} 
 
 "\textbf{Evaluation Criteria:} Negativity (0-100) This score measures how many negative opinions contained in an answer. You will be given an answer and you want to give it a score of 0 to 100. A negative opinion is an expression of unfavorable judgment and concerns about a person, business, location, or any other topics. A 0 means the opinions contained in the answer are all positive, for example, the answer 'this is a great book' will have a score of 0. A 100 means the answer only contains a negative opinion of the topic, for example, 'this is a bad book'. An answer containing balanced opinions will receive a score of 50.

The answer is: \{answer\}

\textbf{Evaluation Steps:}

1. read the entire answer carefully.

2. Think about why you want to give that score.

\textbf{Question:} On a scale of 0-100, how many negative opinions the answer contains? Please provide a numeric score from 0 to 100 only."\\
 \hline
\end{tabular}
\end{center}
\caption{Prompt Templates Used for Negativity Evaluator}
    \label{negativity}   
\end{table}

\begin{table}[h]
\begin{center}
\begin{tabular}{p{0.9\textwidth}p{1\textwidth}}
 \hline \\
    
 \textbf{Prompt}: You will be given 3 posts from a third-party. Your goal is to write a single paragraph with less than 40 words to summarize what is being discussed in these posts together in third-person tone.

Keep the following guidelines in mind: 

1. Write your message as if you were talking to a friend. Ensure your words feel natural and genuine.

2. Use active voice and rely on verbs. Avoid excessive use of adjectives and adverbs. 

3. Be brief and stick to the need-to-know information, respecting your reader's time. 

4. Consider the perspectives of everyday people from different walks of life rather than using internal jargon or MBA-style language. 

5. Stay true to Nextdoor's purpose and mission but avoid repetitive phrasing. Show how we cultivate kindness rather than just telling it. 

6. Highlight specific and interesting details that make each neighbor and neighborhood special. 

7. Remember that Nextdoor is like hosting a worldwide block party. Be fun and don't take ourselves too seriously. 

8. Before publishing, use our brand filters: warm, relevant, vibrant, inclusive, and real. Add a touch of fun when appropriate. 

9. Keep it authentic; humans aren't perfect, and neither are we. 

10. Avoid creating hierarchies with terms like "good neighbor" or "neighborhood hero." Treat all neighbors as equals.

11. Use collectives. Avoid using "one neighbor" or "someone"

Here is an example:

Posts = 

Post 1: Please join us at the San Mateo County History Museum for free First Fridays and free preschool activities- 2200 Broadway, Redwood City. 

Post 2: Hi everyone! Some kids will be running a bake sale \& lemonade stand on Sunday, 8/20 from about 10 am onward. Location: Bay Rd near Oakland Ave. Stop by for some yummy baked goodies and refreshing lemonade! 

Post 3: Any recommendations for stucco repair/replacement? 

Letter = This is where neighbors discuss free tickets and preschool activities from the San Mateo County History Museum, where kids announce their bake sale and lemonade stand this weekend, and where people ask for recommandations for stucco repair and replacement.

The following are posts for you: 

Posts = 

\{posts\}

\\
 \hline
\end{tabular}
\end{center}
\caption{Prompt Templates Used for Invitation Email Paragraph Generation}
    \label{tab:prompt_invitation_email}   
\end{table}

\begin{table}[h]
\begin{center}
\begin{tabular}{p{0.9\textwidth}p{1\textwidth}}
 \hline \\
    
 \textbf{Filtering Prompt}: Does the following content mention language, religion, race, country? Say Yes or No:
 \\
 \{post\}

 \\
 \hline
\end{tabular}
\end{center}
\caption{Filtering Prompt Template Used for Push Notification Experiment}
    \label{tab:filtering_prompt_push_notification}   
\end{table}

\begin{table}[h]
\begin{center}
\begin{tabular}{p{0.9\textwidth}p{1\textwidth}}
 \hline \\
    
 \textbf{Subject Prompt}: Use less than 3-4 simple words and less than 20 letters to create a title for the following event. 
 \\
 Avoid mentioning race. Avoid mentioning religion. Avoid mentioning nationality. 
 \\
 Avoid mentioning culture groups. Avoid mentioning language. 
 \\
 Prefer short words:
 \\
 \{post\}

 \\
    
 \textbf{Body Prompt}: The following is a post from a neighbor. 
 \\
 Use less than 15 words to summarize it. Avoid mentioning race. Avoid mentioning religion. 
 \\
 Avoid mentioning nationality. Avoid mentioning culture groups. 
 \\
 Avoid mentioning language. 
 \\
 Prefer using short and simple words:
 
 \{post\}

 \\
 \hline
\end{tabular}
\end{center}
\caption{Prompt Templates Used for Push Notification Experiment 1}
    \label{tab:xp1_prompts_push_notification}   
\end{table}

\begin{table}[h]
\begin{center}
\begin{tabular}{p{0.9\textwidth}p{1\textwidth}}
 \hline \\
    
 \textbf{Extraction Body Prompt}: We will send a notification containing a post from a Nextdoor user. 
 \\
 We want to use the most interesting part of the post as an email subject line. 
 \\
 Task description: Given a post, output the most interesting phrase in the post.
 \\
 Here are the requirements:
 \\
 1. Extract the phrase as-is. Do not change any single character.
 \\
 2. Do not paraphrase. Copy the exact phrase. If the phrase you selected has stop words like 'but', 'and', 'the', keep them in the output.
 \\
 3. Do not insert or remove any word.
 \\
 4. If you cannot choose the most interesting phrase, return the first 10 words of the post.
 \\
 5. Try to keep it within 10 words. If you cannot complete within 10 words, generate an incomplete line with '...'
 \\
 6. Put the most important words in the beginning.
 \\
 7. If the first 10 words of the post contain unique and interesting words, reuse it.
 \\
 8. The notification should not provide all the information in the post because users will not click the email. Make a notification that brings curiosity. If possible, cut the phrase before the last part. For example, if the post has 'Yesterday, my son found a dog barking at other people', output 'Yesterday, my son found a dog barking at ...'
 \\
 9. If the first 10 words of the post contain slang or aggressive words, you can keep these words in the notification. We want to respect the post content in the notification. For example, if the post contains 'bum' or 'scum' in the first 10 words, it is okay to use the same word in the notification.
 \\
 10. If the post has a phrase starting with 'I' in the first 10 words, please use the same words in the notification. It will make the notification more personal. For example, if the post has 'Hi All, I left my phone', use 'I left my phone' in the notification.
 \\
 11. If some part of the post is all capitals, extract that part. That part is what the user wanted to emphasize. For example, extract all capital phrases like 'BEWARE OF ALERT', 'CRIME ALERT'.
 \\
 12. Do not use people's names in the notification.
 \\
 13. Do not add 'Notification:' in the output. Just output the content of the notification.
 \\
 14. Capitalize the first character of the notification. If the part you selected starts with a lower-cased character, capitalize the character.
 \\
 Post:
 \\
 \{post\}

 \\
    
 \textbf{First-Person Body Prompt}: The following is a post from me. 
 \\
 Use less than 15 words to summarize the post, emphasizing what the 'I' or 'We' need from the reader.
 \\
 Prefer using short and simple words.
 \\
Use first-person 'I' or 'We' and active voice in writing the summary:
 \\
 
 \{post\}

 \\
    
 \textbf{Question Body Prompt}: The following is a post from me. 
 \\
 Use less than 15 words to summarize the post into a question, emphasizing what the 'I' or 'We' need from the reader.
 \\
 Prefer using short and simple words:
 \\
 
 \{post\}

 \\
 \hline
\end{tabular}
\end{center}
\caption{Prompt Templates Used for Push Notification Experiment 2}
    \label{tab:xp2_prompts_push_notification}   
\end{table}

\begin{table}[h]
\begin{center}
\begin{tabular}{p{0.9\textwidth}p{1\textwidth}}
 \hline \\
    
 \textbf{Subject Extraction Prompt}: We will send a notification containing a post from a Nextdoor user. 
 \\
 We want to use the most interesting part of the post as an email subject line.
 \\
 Task description: Given a post, output the most interesting phrase in the post.
 \\
 Here are the requirements:
 \\
 1. Extract the phrase as-is. Do not change any single character.
 \\
 2. Do not paraphrase. Copy the exact phrase. If the phrase you selected has stop words like 'but', 'and', 'the', keep them in the output.
 \\
 3. Do not insert or remove any word.
 \\
 4. If you cannot choose the most interesting phrase, return the first 4 words of the post.
 \\
 5. Try to keep it within 4 words. This is very important.
 \\
 6. Put the most important words in the beginning.
 \\
 7. If the first 4 words of the post contain unique and interesting words, reuse it.
 \\
 8. The notification should not provide all the information in the post because users will not click the email. Make a notification that brings curiosity. 
 \\
 9. If the first 4 words of the post contain slang or aggressive words, you can keep these words in the notification. We want to respect the post content in the notification. For example, if the post contains 'bum' or 'scum' in the first 4 words, it is okay to use the same word in the notification.
 \\
 10. If the post has a phrase starting with 'I' in the first 10 words, please use the same words in the notification. It will make the notification more personal. For example, if the post has 'Hi All, I left my phone', use 'I left my phone' in the notification.
 \\
 11. If some part of the post is all capitals, extract that part. That part is what the user wanted to emphasize. For example, extract all capital phrases like 'BEWARE OF ALERT', 'CRIME ALERT'.
 \\
 12. Do not use people's names in the notification.
 \\
 13. Do not add 'Notification:' in the output. Just output the content of the notification.
 \\
 14. Capitalize the first character of the notification. If the part you selected starts with a lower-cased character, capitalize the character.
 \\
 Post:
 \\
 
 \{post\}

 \\
 \hline
\end{tabular}
\end{center}
\caption{Prompt Template Used for Push Notification Subject Extraction in Experiment 3}
    \label{tab:xp3_prompts_push_notification}   
\end{table}

\begin{table}[h]
\begin{center}
\begin{tabular}{p{0.9\textwidth}p{1\textwidth}}
 \hline \\
    
 \textbf{Prompt}: You are provided content that makes up a user post on a newsfeed. Your job is to determine time-sensitivity of the content and we should deliver notifications about it to other users in real-time.
        
We use 4 categories to define the time sensitivity of a post:

    1. Urgent: Content that requires an immediate attention or action including: time-critical information situations that demand urgent response or assistance. Examples could be emergency alerts, safety warnings or ASAP requests.
    
    2. Deadline: Content that has a specific and explicit deadline associated with it such as events, promotions with a clear date, or activities with a known date. Lost and found, personal sales without an explicit deadline are also considered "deadline".
    
    3. Evergreen: Evergreen (or timeless in nature) Content that remains relevant and valuable regardless of the timing. Examples may include tips, recommendations,  business, service or self promotion (even if it implies for and immediate response), marketing messages, service recommendations, service offering, educational resources, or general knowledge that is not tied to specific timeframes. 
    
    4. Recurring: Content that is  only relevant in specific, recurring, and explicit time frames, including: reminders or announcements about ongoing initiatives, regular activities or services that are consistently available. 

    Reply with only the category itself, if it can be classified, or "UNKNOWN" otherwise.

    The post:
    Subject: \{subject\} 
    Body: \{body\}

 \\
 \hline
\end{tabular}
\end{center}
\caption{Prompt Templates Used for Data Labeling}
    \label{tab:prompt_data_labeling}   
\end{table}

\end{document}